\documentclass[journal]{IEEEtran}

\usepackage{amsmath,amsfonts}
\usepackage[ruled,linesnumbered]{algorithm2e}
\usepackage{array}
\usepackage[caption=false,font=normalsize,labelfont=sf,textfont=sf]{subfig}
\usepackage{textcomp}
\usepackage{stfloats}
\usepackage{url}
\usepackage{verbatim}
\usepackage{graphicx}
\usepackage{cite}
\usepackage{makecell}
\usepackage{xr}

\begin{document}

\title{Cost-driven prunings for iterative solving of constrained routing problem with SRLG-disjoint protection}

\author{P. A. Mosharev, Choon-Meng Lee, Xu Shu, Xiaoshan Zhang, Man-Hong Yung
\thanks{The authors are with Huawei Technologies Co., Ltd., Shenzhen, China }
\thanks{Manuscript received ; revised }

}



\maketitle

\begin{abstract}
The search for the optimal pair of active and protection paths in a network with Shared Risk Link Groups (SRLG) is a challenging but high-value problem in the industry that is inevitable in ensuring reliable connections on the modern Internet. We propose a new approach to solving this problem, with a novel use of statistical analysis of the distribution of paths with respect to their cost, which is an integral part of our innovation. The key idea in our algorithm is to employ iterative updates of cost bounds, allowing efficient pruning of suboptimal paths. This idea drives an efficacious exploration of the search space. We benchmark our algorithms against the state-of-the-art algorithms that exploit the alternative strategy of conflicting links exclusion, showing that our approach has the advantage of finding more feasible connections within a set time limit.
\end{abstract}


\section{Introduction}
\IEEEPARstart{M}{odern} internet services that require reliable and highly predictable low-latency connections range from financial transactions to remote surgery and distributed training of large language models \cite{Spolitis:14}, \cite{Gangidi:24}, motivating development and deployment of new communication standards, such as Deterministic Network (DetNet) \cite{DetNet:19}. Key features of DetNet are the deterministic value of delay and full protection against a single failure of any physical network component.
	
Being deployed at IP level of a network in e.g. IP-over-WDM paradigm \cite{Rajagopalan:00}, \cite{Koo:03}, \cite{Yu:21}, it requires only certain information of underlying physical infrastructure in abstracted forms. Network operation on the IP level is not concerned with bandwidth, wavelength allocation and continuity, or signal-to-noise ratio, rather, the only information from the physical resources that comes to the next layer is the logical link topology, operational cost, signal delay, and Shared Risk Link Groups (SRLGs), which denote any physical component (cable, tube, bridge, etc) shared between several logical links in the network. Since damages to such a component may cause several links to fail simultaneously, the two paths of a single connection must not share any of them, i.e., be SRLG-disjoint.

When formulated as an optimization program, these requirements introduce the SRLG-disjoint Delay Constrained Routing (SRLG-disjoint DCR) problem. Given two points in a network, it is required to find a connection between them. The connection must be as economical as possible in terms of operational cost. To make the connection sustainable against a failure of network components, two separate paths that do not share any SRLG are required. 
Finally, there are also requirements on signal delay along the found paths. Both delays must satisfy the given upper bound, and the difference between them must not exceed the given limit.

The mathematical model of this problem can be seen as a hybrid of two separate problems. The first one, Delay Constrained Routing (DCR) is a particular case of the Constrained Shortest Path problem (CSP), which finds a single path with minimal cost, while satisfying a delay constraint \cite{Joksch:66} - \cite{Thomas:18}. The state-of-the-art for solving the CSP problem is the Pulse algorithm and its variants \cite{Lozano:13} - \cite{Cabrera:20}. The Pulse algorithm is basically a depth-first search with several pruning techniques, enabling a reduction in the search space. The second model is the unconstrained SRLG-disjoint routing problem \cite{Jian:03} - \cite{Xie:18}, whose objective is to find an optimal SRLG-disjoint pair of paths in a network without  constraints. The state-of-the-art algorithm for this problem is CoSE (Conflict SRLG Exclusion).
	
Combination of the two problems gives the
SRLG-disjoint DCR problem. To the best of our knowledge, the only existing algorithm to address directly this problem is CoSE-Pulse+, published in \cite{Zhao:23}. The work \cite{Zhao:23} pointed out, that solution of SRLG-disjoint DCR problem may require, as an intermediate step, to solve a single path DRCR (Delay-Range Constrained Routing) problem with both lower and upper bounds on delay. For this problem it proposed the Pulse+ algorithm, a modification of the Pulse algorithm, to handle the lower bound delay constraint.

However, we make two important remarks, concerning the CoSE-Pulse+ algorithm. First, not all pruning strategies of the original Pulse algorithm are applicable when dealing with lower bound on delay, and not all remaining prunings are equally efficient, as we show below. Second, since the constrained shortest path is not necessarily the shortest one in a network, the search of conflicting SRLG set may not be as effective as the original CoSE algorithm. Besides addressing these issues, our contributions also include the following:
	
\begin{itemize}
  \item We perform statistical analysis of the search space for DRCR and SRLG-disjoint DCR problems and give insights into how to exploit it to speed up the search (Sections \ref{sec:distr_drcr}, \ref{sec:distr_srlg}).
  \item We propose an innovative technique called BTBU, bottom-top bounds update, to improve the performance of Pulse+ algorithm (Section \ref{sec:bottom-top}).
  \item We propose an algorithm for SRLG-disjoint DCR / DRCR problem (Section \ref{sec:corridor}). It makes judicious use of the Pulse / Pulse+ framework while on the high level being intrinsically different from previous CoSE paradigm in a way that it uses updated cost bounds instead of searching for conflicting SRLG sets. We call this algorithm BTCS, bottom-top corridor search.
  \item We extend the dataset from \cite{Zhao:23} by adding a set of scale-free networks for better coverage of different network topologies (Section \ref{sec:experiment} and Appendix \ref{app:dataset}).
\end{itemize}
	
We benchmark our algorithms on the wide range of networks of different type and size, providing a thorough comparison with Pulse+ and CoSE-Pulse+ algorithms and highlighting their advantages and disadvantages (Section \ref{sec:experiment}). From the benchmark results and analysis, we conclude that our algorithms are better suited to the case when one needs to find as many feasible solutions as possible, meeting strict time requirements. On the other hand, CoSE-Pulse+ has the advantage in proving the non-existence of feasible solutions and better worst-case performance in certain scenarios.

\section{Problem statement and background}
	
In this section we give a mathematical description for the problems to be solved and an overview of the Pulse algorithm.

Let $G(N, E)$ be a graph, representing the network. Here, $N$ is a set of nodes, $E$ is a set of edges. Each edge $e$ of the graph has two positive real-valued parameters: cost, denoted by $c(e)$, and delay, denoted by $d(e)$. A path $P$ is a sequence of edges. The sum of costs of all edges along a path is denoted as $c(P)$, and sum of delays as $d(P)$. Edges of the graph are grouped into SRLGs i.e an SRLG is a set of one or more edges. An edge can in principle belong to several SRLGs.

\subsection{DRCR problem}
Given a source node $s$, target node $t$, delay lower bound $d_{low}$ and delay upper bound $d_{up}$, find the optimal path $P_{s \rightarrow t}$, connecting $s$ to $t$, and satisfying constraints: \\
		
	\fbox{\begin{minipage}{23em}
		\begin{align*}
			\min\limits_{P} \; &c(P_{s \rightarrow t}) \\
			\mbox{\textbf{s. t.}} \;	&P \mbox{ is any elementary path from } s \mbox{ to } t, \\
			&d_{low} \le d(P_{s \rightarrow t}) \le d_{up}
		\end{align*}
	\end{minipage}}
	\\
	
Here and thereafter, since cost is the objective value to be minimized, we will use words "short" and "long" referring to the cost. For example, shortest path is the path with the minimal cost.

We present in-depth analysis of this problem in Section \ref{sec:distr_drcr} and propose a new algorithm to solve it in Section \ref{sec:bottom-top}.
	
\subsection{SRLG-disjoint constrained routing problem}
	\label{sec:SRLG_problem}	
	
	Given source node $s$, target node $t$, delay lower bound $d_{low}$, delay upper bound $d_{up}$ and delay-difference limit $d_{diff}$, find a pair of SRLG-disjoint paths $AP$, $PP$, both connecting $s$ to $t$ and satisfying the delay range and delay diff constraints, and $AP$ has the minimal cost: \\
	
	\fbox{\begin{minipage}{23em}
		\begin{align*}
			\min\limits_{AP, PP} \; &c(AP) \\
			\mbox{\textbf{s. t.}} \;	&AP, PP \mbox{ are two elementary paths from } s \mbox{ to } t,\\
			&d_{low} \le d(AP) \le d_{up} \\
			&d_{low} \le d(PP) \le d_{up} \\
			&\left|d(AP) - d(PP)\right| \le d_{diff} \\
			&AP \mbox{ and } PP \mbox{ do not share any SRLGs.}
		\end{align*}
	\end{minipage}}
	\\
	
Here, $AP$ and $PP$ stands for Active Path and Protection Path. In a typical scenario, all traffic is routed by AP, while PP is used only in the case of AP failure. So, it is reasonable to minimize cost only for AP, letting PP be any available path. We note that this is not the only possible choice for the objective. Another important example is so called min-sum problem, that requires to minimize total cost of the two disjoint paths. This class of problems is typically approached via modifications of Suurballe's heuristic \cite{Suurballe:74}, \cite{Todimala:04}. Our problem thus classifies as min-min problem.

The problem with a nontrivial lower bound on delay may appear in the following scenario. Different internet services have different requirements for connection quality \cite{DiffServ:06}. There are tasks mentioned in the Introduction that rely on limited and deterministic delay values. Other applications, such as distributed cloud office, can tolerate tens of milliseconds in delay without quality degradation, while others, like e-mail or file transfer, have pretty much no requirements for signal delay. When all these services occupy the same infrastructure, it may be reasonable to set an artificial lower bound on delay for non-critical applications to reserve low-latency connections for more valuable tasks.
	
We note that in case $d_{low} = 0$ the problem reduces to SRLG-disjoint DCR problem that is more widely applicable and better studied in literature. In particular, CoSE-Pulse+ algorithm designed in \cite{Zhao:23} handles only  $d_{up}$ constraint.
	
All SRLG-disjoint routing problems can be classified into non-trap and trapped instances, the latter being more challenging to solve. The notion of trap originates from a solution technique called Active Path First, or APF \cite{Dunn:94}, \cite{Liu:01}, \cite{Xu:02}. In this technique, the DRCR problem for AP is first solved, then all edges sharing SRLG with the found AP are removed from the graph, and delay bounds are updated according to $d_{diff}$. The DRCR problem is then solved again in the graph with removed edges to find PP. If it succeeds, the solution satisfies all the requirements and is optimal. But in some cases the shortest candidate for AP does not have feasible disjoint pair, that could serve as PP. This is the so-called trap in the literature, and is of particular interest for research, since in this case simple APF approach is not enough. Traps can be avoidable and unavoidable, depending on whether or not a feasible pair of paths exists in principle. For the case of avoidable trap, some sophisticated strategy is required to find the solution \cite{Xu:03}. For unavoidable traps, sometimes proof of infeasibility is needed.

We analyze the search space of SRLG-disjoint DCR problem in Section \ref{sec:distr_srlg} and propose a novel algorithm to solve it in Section \ref{sec:corridor}.

We also note that in case of unconstrained problem and planar network topology polynomial-time algorithms are already known for certain types of SRLG distribution, see for example \cite{Vass:22}, \cite{Berczi-Kovac:24}. In current work we consider constrained problems, and do not leverage any particular network topology. On the contrary, in Section \ref{sec:experiment} we test our algorithms on as much diverse dataset as we can, to highlight all aspects of its performance.

\subsection{Pulse algorithm}
\label{sec:Pulse}

In this section we briefly describe the Pulse / Pulse+ family of algorithms and how they are incorporated into our work.
	
The original Pulse algorithm \cite{Lozano:13} was designed to solve the CSP problem with only the upper-bound delay constraint. The key idea in the Pulse algorithm is similar to the branch-and-bound approach from the integer programming literature. The algorithm introduces variable $current\_min\_cost$ to store the cost of the shortest path found so far. In the beginning, this variable is set to be infinity. In the initialization stage, the tree of shortest paths from each node to the target node is found, separately with minimal cost and delay. So, the minimal possible cost and minimal possible delay of a path from any node $u$ to target node is known. We denote them $c(P_{u \rightarrow t}^{min\_cost \ })$ and $d(P_{u \rightarrow t}^{min\_delay})$, respectively. This search can be efficiently done with Dijkstra or A* algorithm. 
	
In the main part of the algorithm, pulses propagate through the graph following the depth-first search strategy. To reduce the search space, three pruning strategies are implemented. When a pulse reaches some node $u$, the algorithm checks it for infeasibility, optimality and dominance. 
	
Infeasibility pruning cuts a branch, if
	
\hspace{\parindent} $d(P_{s \rightarrow u}) + d(P_{u \rightarrow t}^{min\_delay}) > d_{up}$.
	
Optimality \ pruning cuts a branch, if
	
\hspace{\parindent} $c(P_{s \rightarrow u}) + c(P_{u \rightarrow t}^{min\_cost \ }) \ge current\_min\_cost$.

Dominance pruning cuts a branch, if there is information stored at this node, that previously a pulse with lower delay and lower cost already reached this node.
	
The successful implementation of Pulse algorithm has 3 main components: search order, pruning strategies and the action on the terminal node. Some attempts to improve search order were made in works \cite{Bolivar:14}, \cite{Cabrera:20}, \cite{Zhao:23} to further increase performance of the original Pulse for the CSP problem. \cite{Zhao:23} also proposed a change in pruning strategies to handle lower bounds on delay, producing the Pulse+ version of the algorithm. It differs from the original Pulse in two points: firstly, the dominance pruning is not applicable in this case, and secondly, tracking of visited nodes is required for each pulse to avoid cycles in the found path. The pseudocode of Pulse+ algorithm is presented in Algorithms \ref{alg:Pulse+_high} and \ref{alg:Pulse+_recursive}.
	
\begin{algorithm}
		
		\SetKwInOut{Input}{input}\SetKwInOut{Output}{output}
		\SetAlgoLined
		\Input{Graph G, source $s$, target $t$, delay range $[d_{low}, d_{up}]$.}
		\Output{The shortest path from $s$ to $t$, satisfying delay range constraint}
				
		$P = \{\}$;  $P^* = \{\}$;  $visited\_nodes = \{\}$
		
		$current\_cost = 0$;    $current\_delay = 0$
		
		Initialization(G, $t$).  /* \textit{calculate min cost and min delay trees from each node to target with the help of Dijkstra or A* algorithm} */
		
		$current\_min\_cost = init\_value$  /* \textit{infinity for original Pulse+ or upper cost bound in BTBU} */
		
		Pulse\_plus($P^*$, $P$, $s$, $visited\_nodes$, $current\_cost$, $current\_delay$, $current\_min\_cost$)
		
		\Return current\_min\_cost, $P^*$
		
		\caption{Pulse+ algorithm}
		\label{alg:Pulse+_high}
	\end{algorithm}

	\begin{algorithm}
		
		\SetKwInOut{Input}{input}\SetKwInOut{Output}{output}
		\SetAlgoLined
		\Input{current optimal path $P^*$, current path $P$, current node $n$, list of visited nodes $visited\_nodes$, current cost $c$, current delay $d$, $current\_min\_cost$}
		\Output{void}
		
		\If{$n == t$ \textbf{and} $d_{low} \le d \le d_{up}$ \textbf{and} $c < current\_min\_cost$}{
			$P^* = P$
			
			$current\_min\_cost = c$
			
			\Return
		}
		
		\For{$(nk)$ in egress links of node $n$}{
			\lIf{$k$ \textbf{in} $visited\_nodes$}{continue}
			\lIf{$d + d(nk) + d(P^{min\_delay}_{k \rightarrow t}) > d_{up}$}{\label{alg:pulse+:infeas}continue}
			\lIf{$c + c(nk) + c(P^{min\_cost}_{k \rightarrow t}) > current\_min\_cost$}{\label{alg:pulse+:opt}continue}

			Pulse\_plus($P^*$, $P + (nk)$, $k$, $visited\_nodes + \{k\}$, $current\_cost + c(nk)$, $current\_delay + d(nk)$, $current\_min\_cost$)
			
		}
		\caption{Pulse\_plus()  recursive function}
		\label{alg:Pulse+_recursive}
	\end{algorithm} 
	
One weakness of Pulse / Pulse+ algorithm originates from the fact that optimality pruning comes into play only after the first feasible path was found. So, there is no pruning by cost for the search of the first feasible path. As shown in the next section, this first-found feasible solution is often far from optimality due to statistical reasons, because the majority of paths are situated at higher values of cost. In section \ref{sec:bottom-top} we address this problem by introducing iteratively updated heuristic upper-bounds on cost. This enables pruning by cost from the start of the search, thus effectively restricting the search space.
	
In section \ref{sec:corridor} we further develop this idea and provide an efficient technique for the solution of SRLG-disjoint DRCR problem, that uses modifications of Pulse+ as a subroutine. For that, we will modify the third component of the algorithm, namely the action at a terminal node. Thus we obtain three versions of the algorithm: the original Pulse+ searching for the shortest single constrained path, also an algorithm to search for all AP candidates in given cost range and another one to search for a feasible PP without optimality requirement. The pseudocode and discussion of these versions can be found in Appendix \ref{app:pulse+}.

\section{Algorithms}
\label{sec:Algorithms}
	
As mentioned in the previous section, only two pruning strategies are applicable in the presence of a delay constraint: infeasibility and optimality pruning. While the first is deterministic, the second depends on the cost of the first found feasible path and how fast it is found. Our innovation aims to use pruning by cost in a more controllable way. We can gain insight into this by analyzing the structure of search space with respect to the path cost.

\subsection{Path distribution for DRCR problem}
	\label{sec:distr_drcr}
\begin{figure*}[!t]
		\centering
		\includegraphics[width=1\textwidth]{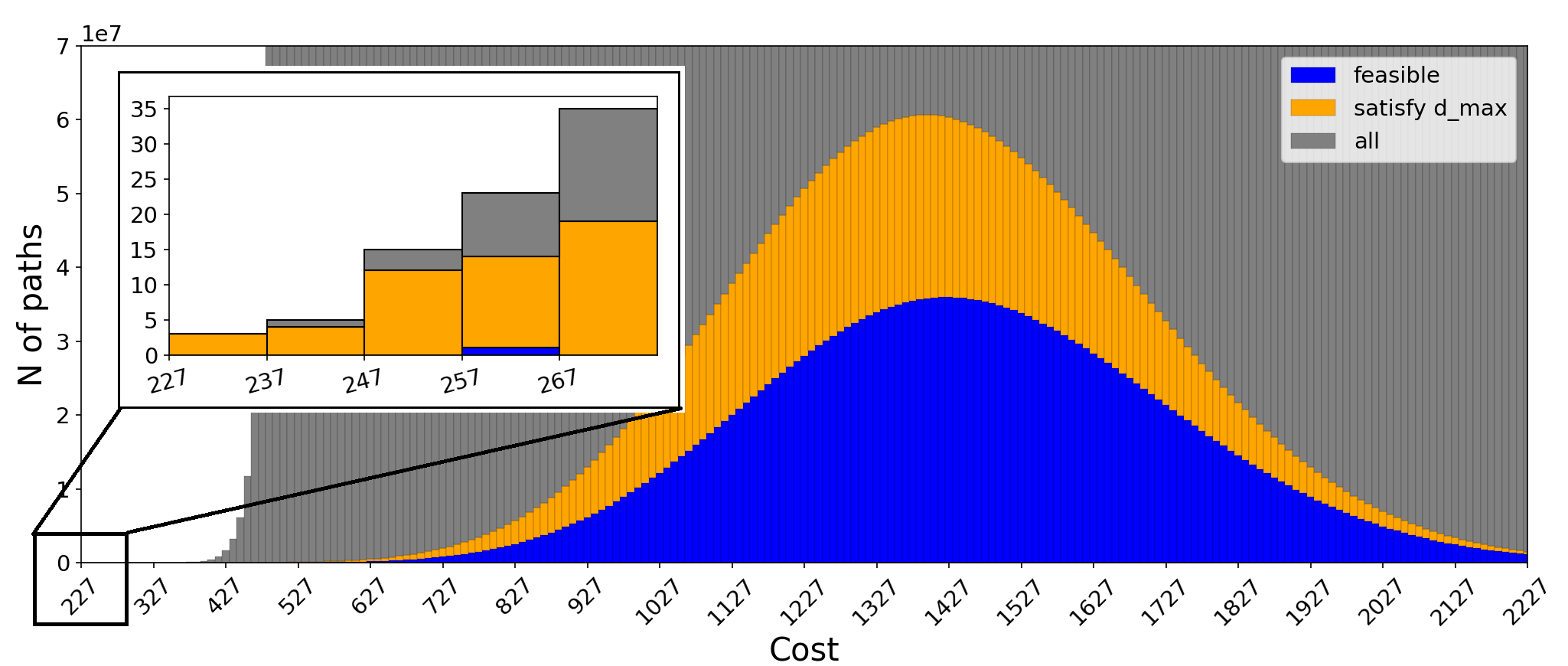}
		\caption{Search space structure for a typical instance of DRCR problem on a graph with 1000 nodes and 7050 links.  
		}
		\label{fig:CSP_distribution}
	\end{figure*}

To illustrate the intuition behind our algorithms, we describe the distribution of paths for a typical problem instance (Figure \ref{fig:CSP_distribution}). This example is qualitatively representative for many problem cases. We provide more examples for other network topologies in the Appendix \ref{app:tails}. On the figure the path distribution relative to their cost for one particular task is shown. The data for this diagram is obtained in two steps. First, ordinary Pulse+ algorithm without optimality pruning is adopted to get the distribution of feasible paths. It takes about 40 minutes to finish. To count all paths including infeasible we use a modification of Pulse+ algorithm that is designed for all path search in cost corridors (see Appendix \ref{app:pulse+all}) with relaxed delay constraint and termination when path count reaches $10^8$. It takes few hours to complete.
	
\begin{figure*}[!t]
		\centering
		\includegraphics[width= 1 \textwidth]{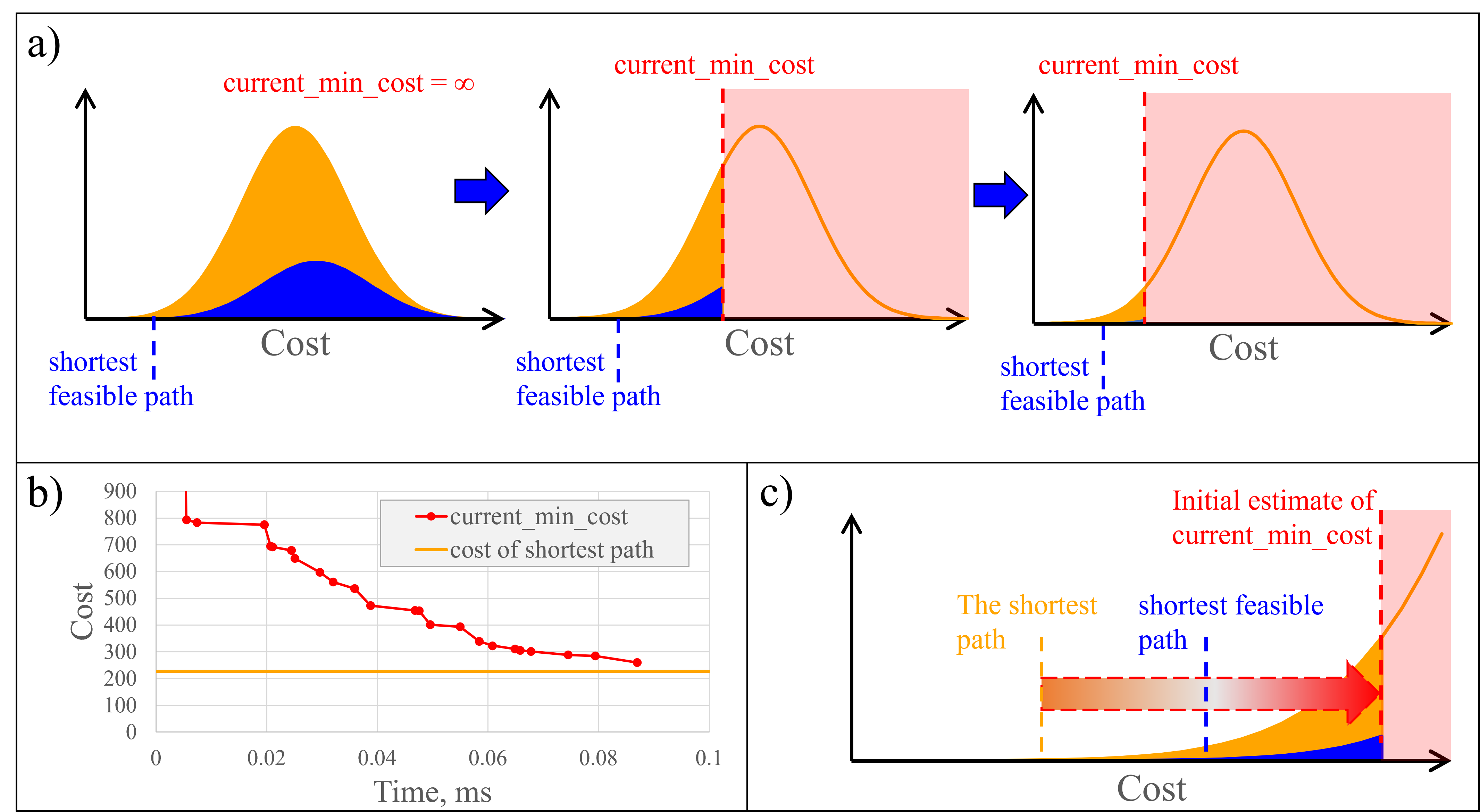}
		\caption{Dynamics of $current\_min\_cost$ in the Pulse algorithm. a) The distribution of feasible paths (blue) and search space that is left after infeasibility pruning (orange) in the process of Pulse algorithm updates. b) Actual cost dynamics for one typical task. c) Our proposal to estimate initial value of $current\_min\_cost$ based on the cost of the shortest unconstrained path}
		\label{fig:CSP_dynamics}
	\end{figure*}

	The diagram shows that paths satisfying the $d_{up}$ constraint follow a bell-shaped distribution around some pretty high cost value. The total number of paths in this category is more than 4 billion, which is, in effect, the size of the initial search space. At the beginning, $current\_min\_cost$ is set to infinity, so the search is conducted among all paths that satisfy $d_{up}$ constraint (Figure \ref{fig:CSP_dynamics} (a)). Since the DFS-like search in the original algorithm is unguided (basically, random), the most probable path to be found tends to lie in the region where feasible paths are most frequent. After that, the optimality pruning comes into play, installing the upper bound on cost and discarding the right-hand part of path distribution. However, due to the bell-like shape of the distribution, more frequent thus more probable to find paths still incline towards the higher values of cost. So, each next path to be found will most probably have cost value close to the current value of $current\_min\_cost$. Consequently, the value of $current\_min\_cost$ will be decreased only a little on each iteration, and the search process will resemble slow motion "down the hill" by the feasible path distribution.
	
	The Figure \ref{fig:CSP_dynamics} (b) shows how the $current\_min\_cost$ value is updated during the search process in original Pulse+ case. At the beginning it is set to infinity. The first feasible path can be found readily, but it has relatively high value of cost, which then slowly reduces towards the minimum value. The cost of the shortest feasible path appears to be very close to the cost of the shortest unconstrained path. So an intuitive strategy for estimating the initial cost is to set it close to the solution of the unconstrained problem, which is known from initialization stage.

\subsection{Bottom-top exploration approach to DRCR problem}
	\label{sec:bottom-top}
	For the single-path CSP problem, our work is a natural extension of the Pulse+ algorithm, which we call Bottom-Top Bounds Update (BTBU). Recall that the initialization step of the Pulse / Pulse+ algorithm calculates the shortest path tree from the target node to each node in the graph. So, before the Pulse algorithm starts, we have a reference lower-bound value for cost, which is the cost of the shortest path from the source to the target. The approach of Pulse+ is to set $current\_min\_cost=infinity$ and run the algorithm. Instead, we propose to place a judiciously chosen upper bound on the cost and search for feasible paths under this upper bound. If no such paths are found, the chosen upper bound is too tight and lies below the cost of the shortest feasible path. In this case, the upper bound is updated according to a specific strategy and the search is repeated again until a feasible path is found. At first glance, it may seem ineffective because many re-runs may be needed before finding the feasible solution. Nevertheless, all unsuccessful runs will be performed near the tail of the path distribution between the cost of the shortest unconstrained path and the cost of the shortest feasible path, meaning that very little time is spent exploring nearly empty search space. The initial estimate of the upper bound and the update rule for it must be chosen so that the upper bound at each iteration is not too loose to effectively restrict the search space, while at the same time keeping number of necessary algorithm reruns reasonably low before the feasible region is reached.
	
	\begin{algorithm}
		
		\SetKwInOut{Input}{input}\SetKwInOut{Output}{output}
		\SetAlgoLined
		\Input{Graph G, source $s$, target $t$, delay range $[d_{low}, d_{up}]$, cost of the shortest path from $s$ to $t$: $shortest\_cost$}
		\Output{The shortest path from $s$ to $t$, satisfying delay range constraint}
		
		$current\_min\_cost = shortest\_cost$
		
		\While{Path not found}{
			
			$current\_min\_cost \ *= \ 2$
			
			Pulse+(G, $s$, $t$, $d_{low}$, $d_{up}$, $current\_min\_cost$)
			
		} 
		\Return{Path, satisfying the constraints}
		\caption{Bottom-top bounds update for Pulse+ (doubling cost bound)}
		\label{alg:Pulse++1}
	\end{algorithm}
	
	\begin{algorithm}
		\SetKwInOut{Input}{input}\SetKwInOut{Output}{output}
		\SetAlgoLined
		\Input{Graph G, source $s$, target $t$, delay range $[d_{low}, d_{up}]$, cost of the shortest path from $s$ to $t$: $shortest\_cost$}
		\Output{The shortest path from $s$ to $t$, satisfying delay range constraint}
		
		$cost\_step = f(G)$   /* \textit{choose cost step based on costs of edges in the graph, see discussion in the text} */
		
		$current\_min\_cost = shortest\_cost + cost\_step$
		
		\While{Path not found}{
			Pulse+(G, $s$, $t$, $d_{low}$, $d_{up}$, $current\_min\_cost$)
			
			$cost\_step \ *= \ 2$		
			
			$current\_min\_cost \ += \ cost\_step$
			
		} 
		\Return{Path, satisfying the constraints}
		\caption{Bottom-top bounds update for Pulse+ (doubling step)}
		\label{alg:Pulse++2}
	\end{algorithm}	
	
	Two approaches were found to be effective in our experiments (Algorithms \ref{alg:Pulse++1}, \ref{alg:Pulse++2}). For the first one, the initial value of current\_min\_cost is two times larger than the cost of the shortest unconstrained path. On each iteration, if the search was unsuccessful, the cost is again multiplied by 2. For the second approach, we introduce a variable called $cost\_step$ which is given by some function of the cost of a graph edge e.g. the cost of the shortest edge, or average cost of an edge in the graph, or some other estimation. The search starts from value of $current\_min\_cost$ just one single step higher than the cost of the shortest unconstrained path. When an iteration is unsuccessful, the step is multiplied by 2, added to the $current\_min\_cost$ and the Pulse+ search runs again.

In both algorithms the step is multiplied by 2 on each iteration, thus the bound increases in geometric progression. Such update rule was found the most effective in our experiments. Algorithm \ref{alg:Pulse++1} is more minimalistic, since both the initial estimate for the upper bound and the update rule are based solely on the cost of the shortest unconstrained path. Algorithm \ref{alg:Pulse++2} allows to choose proper function $f(G)$, that may better match the particular network. In our experiments we used the cost of the shortest edge in the graph for the $f(G)$, or double that value.

Section \ref{sec:experiment} will give a performance comparison between the two algorithms and the original Pulse+ algorithm.

\subsection{Path distribution for SRLG-disjoint DCR problem}
\label{sec:distr_srlg}

\begin{figure*}[!ht]
	\centering
	\includegraphics[width=1 \textwidth]{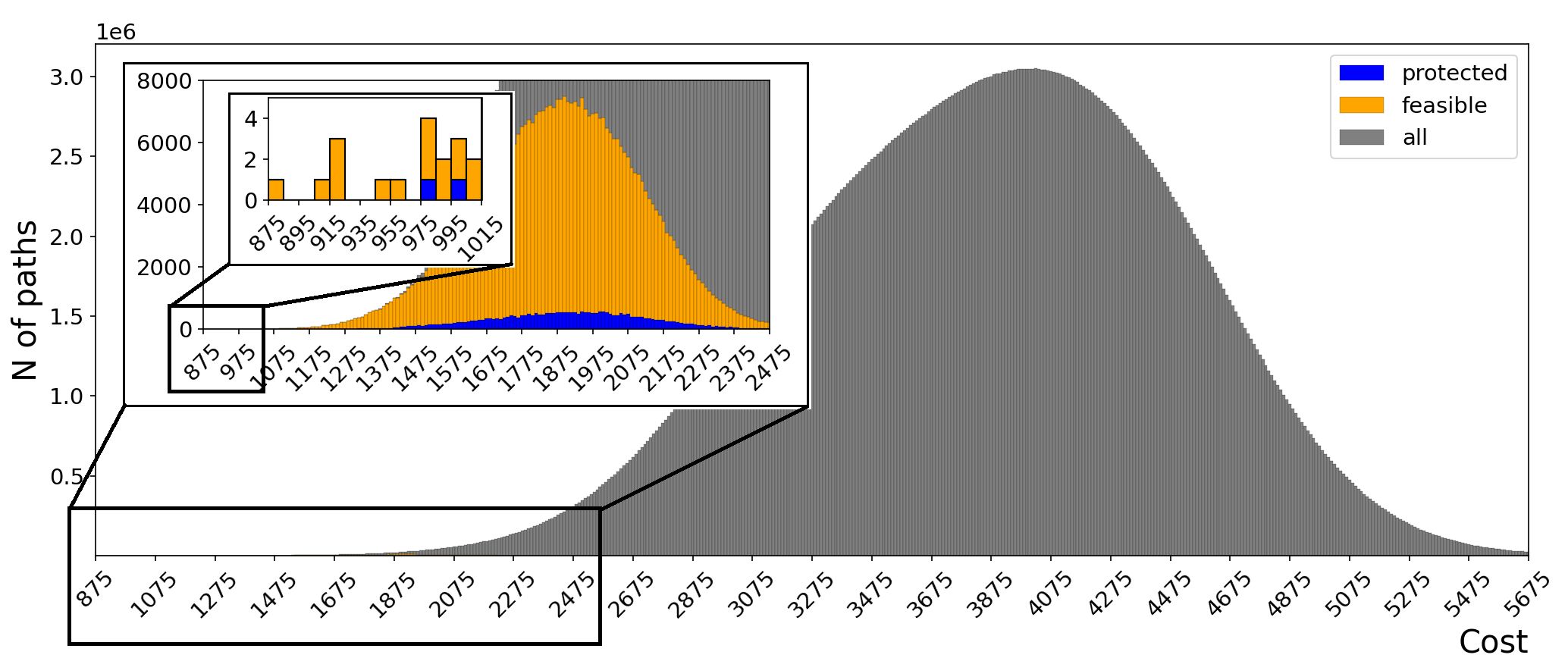}
	\caption{Distribution of active path candidates for SRLG-disjoint DCR problem. The graph instance used to obtain this data contains 197 nodes and 490 links. Feasible paths are those satisfying delay constraint, protected are those satisfying delay constraint and having an SRLG-disjoint protection path that satisfies delay constraints.}
	
	\label{fig:SRLG_space}
\end{figure*}	

	In this section we describe the intuition behind our BTCS algorithm for the SRLG-disjoint D(R)CR problem. Figure \ref{fig:SRLG_space} shows the distribution of active paths for a typical instance of the problem. The network, used to generate this data is smaller than the one chosen for section \ref{sec:distr_drcr} (only 197 nodes and 490 links), so it is possible to plot the distribution of all paths from source to target for this task. We put the partial distributions for several other instances into Appendix \ref{app:tails} to show, that this picture in its essential features is indeed typical for the problem under consideration.
	Although the distribution of paths on Figure \ref{fig:SRLG_space} may look similar to that on Figure \ref{fig:CSP_distribution}, note that the legend of two figures is different.
	
	Recall the problem statement for SRLG-disjoint DCR problem from section \ref{sec:SRLG_problem}. We are interested not only in the minimum cost of the active path (AP) but also in one that satisfies delay constraints and has a protection path (PP). Let us look at the problem from the perspective of AP while considering the existence of a disjoint PP as a property of each AP. We can divide all paths that connect the given source node to the target node into three categories. The first category contains infeasible paths that do not satisfy delay constraints. The second category has paths that satisfy constraints but do not have a disjoint PP. The third category has paths that satisfy constraints and have at least one disjoint PP. The shortest path in the last category and the corresponding disjoint PP are the paths we are looking for. We call such a pair of shortest feasible AP together with disjoint PP the solution of the problem.
		
At the left end of the distribution we can notice, that the shortest (lowest cost) candidate for the active path does not have disjoint protection. This is the previously discussed trap situation. Contrary to some rather exaggerated examples presented in literature (see, for example, \cite{Zhao:23} and discussion in Appendix \ref{app:tails}), we notice that the shortest path having protection pair is also located close to the shortest one, near the left tail of the distribution. Therefore the natural idea is again to restrict the search in the area of low costs, instead of searching across the whole space of feasible paths.

\subsection{Solving SRLG-disjoint DRCR problem: corridor search}
	\label{sec:corridor}
	
\begin{figure*}[!ht]
		\centering
		\includegraphics[width=1 \textwidth]{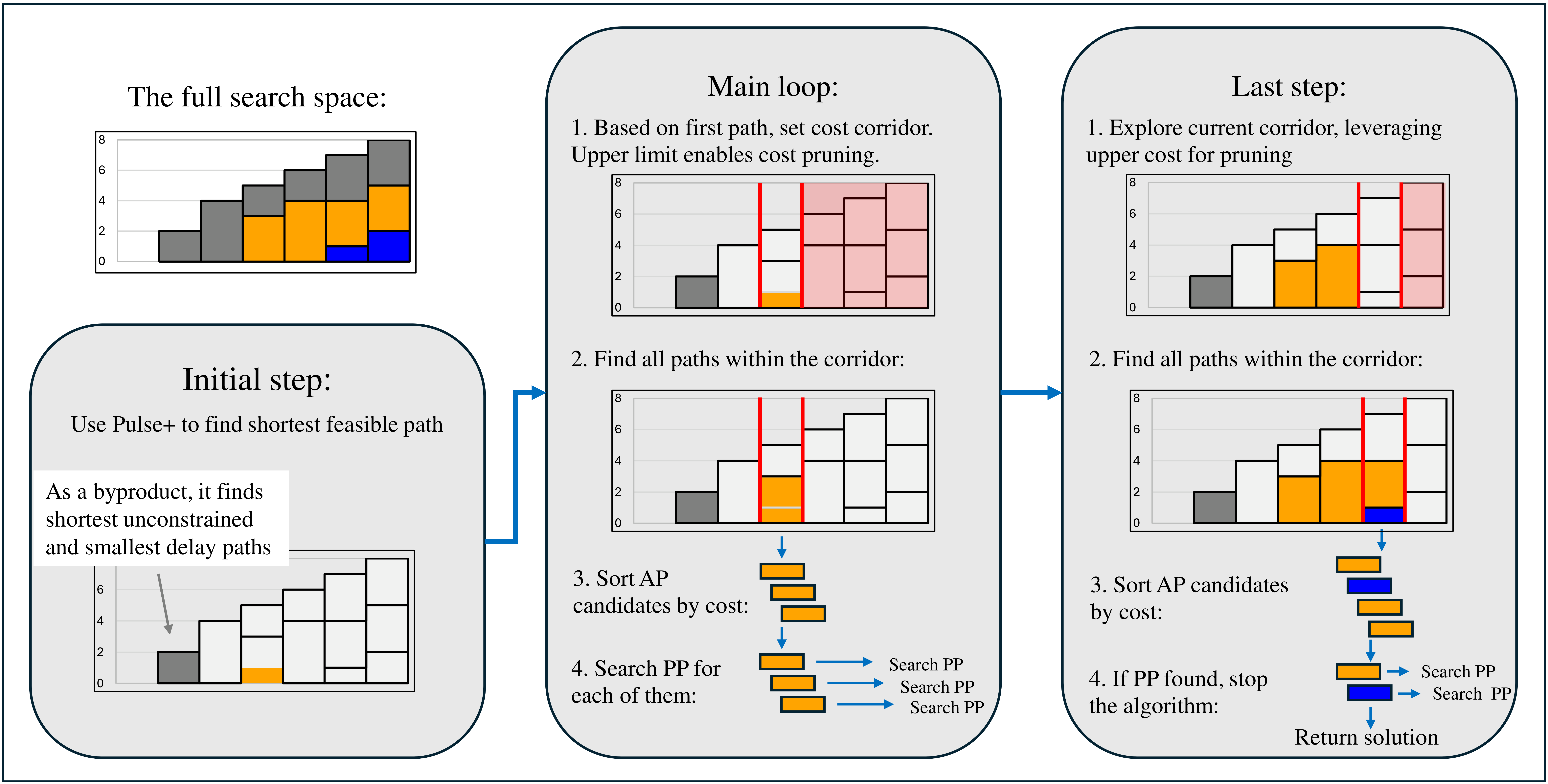}
		\caption{Graphical illustration of BTCS algorithm}
		
		\label{fig:BTCS_description}
	\end{figure*}

	Here we describe our proposed algorithm, that uses several adaptations of Pulse+ as intermediate steps. We call it BTCS-Pulse+, bottom-top corridor search. BTCS-Pulse+ is exact in the sense that it guarantees optimality of the found solution. In terms of performance, it has both advantages and disadvantages compared to COSE-Pulse+ (see discussion in Section \ref{sec:experiment}). The main features of the technique are depicted on Figure \ref{fig:BTCS_description} and the pseudocode is presented in Algorithm \ref{alg:SRLG}.
	
	The initial stage of the algorithm consists of a single run of Pulse+. It efficiently handles the delay constraints and returns the shortest feasible AP candidate. It cannot account for the existence of a disjoint PP for this candidate, which has to be tested separately. For that, the algorithm removes from the graph all edges that share an SRLG with the found AP and updates the delay lower and upper bounds according to the delay difference constraint based on the delay of the found AP. Removal of the links that share SRLG with AP guarantees that AP and PP will not share any SRLGs. And update of the upper and lower bounds of delay is necessary to ensure that the difference in delay between AP and PP satisfies the delay difference constraint. After this, another run of Pulse+ is performed on the modified graph with new delay range constraint to find a PP. If PP is successfully found, then this pair of AP and PP is the solution of the problem, and algorithm stops. 
	
	If this first AP candidate does not have a PP, the algorithm starts to search for other candidates. Since we are interested in finding the shortest protected AP, it is natural to start from the already found one which has the lowest cost and gradually move to the paths of higher costs, searching for feasible AP candidates and checking them one by one for the existence of disjoint PP, until such candidate is found. 
	
	The cost of the initially found shortest AP candidate serves as a starting point. The whole range of costs above it is divided into a sequence of reasonably small corridors. The corridor width is a hyperparameter of the algorithm. It can be estimated based on the costs of the graph edges, tuned by hand or pretrained on historic data for the same network. The algorithm will explore those corridors one by one, starting from the smallest cost and moving in the bottom-top direction.

The main loop of the algorithm explores one particular cost corridor, that is given by current lower and upper bounds on cost. To efficiently find all feasible AP candidates in a corridor, we propose a modification of the Pulse+ algorithm, that uses the upper side of the corridor to prune the search space above, and enumerates all feasible paths within the corridor instead of searching for the shortest one. We put the details of this Pulse+ implementation in Appendix \ref{app:pulse+all}. This version of the Pulse+ algorithm uses cost and delay in a similar way, as delay range is given in the problem statement, and cost range corresponds to the corridor. The upper bound of the cost corridor serves for pruning by cost together with the original infeasibility pruning by the upper bound on delay. And instead of updating the cost bound, as it is done in the original Pulse+ to find the optimal path, this version store all paths, that satisfy upper and lower bounds on delay and cost. Thanks to the pruning by both delay and cost, the effective search space is kept to be sufficiently small (it corresponds to the left tail of the feasible distribution on Figure \ref{fig:SRLG_space}). If the corridor width is chosen to be small enough, the number of AP candidates found in each corridor is also reasonably low.
	
The AP candidates found in current cost corridor are sorted by cost in ascending order and taken one by one to search for a disjoint PP for each of them. It is done in a similar manner to what was described above for the first found AP candidate: removing from the graph all edges, that share SRLG with current AP candidate, updating upper and lower bounds on delay based on the delay of current AP and delay difference constraint, then using Pulse+ to search for PP. If a disjoint PP is found for one of the AP candidates, then this pair of paths is the solution for the problem, and algorithm stops. If there is no disjoint PP for all AP candidates in the current corridor, then the algorithm proceeds to the next corridor by updating the lower and upper bounds on cost and looks for AP candidates again.
	
This procedure is repeated until the solution is found. Since in a typical case the AP, that has a disjoint pair, is situated close to the lower end of the distribution (See Fig. \ref{fig:SRLG_space}, Appendix \ref{app:tails}, and also Fig. \ref{fig:SRLG_tasks} (d), (e) and corresponding figures in Appendix \ref{app:worst_case}), this search typically requires only a few corridors that contain only a few feasible AP candidates, which makes the search fast and efficient. Take for example the task on Fig \ref{fig:SRLG_space}. If the width of the cost corridor is equal to $10$, then each corridor corresponds to a single bin of the diagram. In this case, $11$ corridors will be explored, before the shortest AP that has a disjoint PP is found (the one shown in blue in the corridor between cost values $975$ and $985$). Five of those corridors are empty, and total number of AP candidates necessary to check (the yellow ones having costs lower, than the first blue) will be slightly above $10$. 
	
\begin{algorithm}
	\SetKwInOut{Input}{input}\SetKwInOut{Output}{output}
	\SetAlgoLined
	\Input{Graph G, source $s$, target $t$, delay range $[d_{low}, d_{up}]$, delay difference $d_{diff}$}
	\Output{The pair of SRLG-disjoint paths from $s$ to $t$, satisfying delay range constraint and within delay difference from each other; active path is shortest possible}
	
	calculate $starting\_cost$ and first AP candidate by Pulse+ algorithm /* \textit{See Section \ref{sec:Pulse}} */ \label{algSRLG:line:1}

	remove edges sharing SRLG with AP from the graph 
	
	use Pulse+ to find PP, satisfying constraints  /* \textit{See Appendix \ref{app:pulse+feas}} */	\label{algSRLG:line:2}
	
	\If{PP not found}{
	
	$cost\_step = f(G)$ /* \textit{Compute the corridor width based on the costs of the edges in the graph or choose it as a hyperparameter} */
	
	$c_{low} = starting\_cost$
	
	$c_{up} = starting\_cost + cost\_step$
	
	\While{PP not found}{
		
		use Pulse+ to search for all AP candidates with $c_{low} \le c(AP) < c_{up}$, $d_{low} \le d(AP) \le d_{up}$. /* \textit{See Appendix \ref{app:pulse+all}} */  \label{algSRLG:line:all}
		
		sort AP candidates in ascending cost order
		
		\For{each AP candidate}{
			remove edges, sharing SRLG with AP
			
			\If{check\_if\_graph\_is\_connected(G, s, t)} 
			{
				\label{algSRLG:line:connected}
				use Pulse+ to search for PP, satisfying constraints /* \textit{See Appendix \ref{app:pulse+feas}} */
				\label{algSRLG:line:PP_search}
			
				\If{PP found}{return AP and PP}
			}		
		}
		
		$c_{low} \ += \ cost\_step$
	
		$c_{up} \ += \ cost\_step$
		
		} 
	\Return{"no feasible solution"}
	}
\caption{Corridor search for delay-constrained SRLG-disjoint pair}\label{alg:SRLG}
\end{algorithm}	

The above description outlines the algorithm in all its distinctive features. Now let us discuss some improvements, that make it even more efficient and robust.

First of all, note, that the problem statement for min-min SRLG-disjoint DRCR problem does not require PP to have optimal cost. So, there is no need to use the full version of Pulse+ in the search for PP. Instead we propose a simplified version of Pulse+, that returns the first found feasible protection path. It still performs the infeasibility pruning by delay and guarantees the satisfaction of constraints, but it does not track the cost of the found path and does not update the found solution to minimize the cost. More details of it can be found in the Appendix \ref{app:pulse+feas}.

We also noticed that in many cases, the removal of links that share SRLGs with the current AP candidate leads to the source and target nodes appearing in disconnected clusters. Therefore, we perform a simple connectivity check between the source and target node after removing edges (lines 13 and 14 in Algorithm 3). A simple exhaustive DFS search can do this connectivity check without worrying about the constraints. This search has the complexity of $O(|V|)$, where $|V|$ is the number of nodes and thus is very effective.

BTCS allows parallel execution that can be implemented on two different levels. Firstly, check for protection paths for each found AP in given corridor can be done in parallel. Also, since the search in each subsequent cost corridor does not depend on the results from previous corridors, several algorithm "steps" can be performed simultaneously in parallel in several corridors. For the current paper, we implemented only the second option since its efficiency is closely connected to the distribution of AP candidates, which gives further insight into the proposed technique. In section \ref{sec:experiment}, we show experimental results for the algorithm, implemented in one, two, and four threads, and discuss it.

\section{Experiment}
	\label{sec:experiment}
	
	\subsection{Dataset and metrics}
	
We evaluate performance of our algorithms, using dataset that contains networks published with the paper \cite{Zhao:23} (check the link \cite{Drcr_Github}), and a set of scale-free networks generated by ourselves.
 
The dataset \cite{Drcr_Github} contains network topologies of two types: real-world and randomly generated networks. Real-world networks originate from the Topology Zoo collection \cite{Zoo}. Random networks are Erdos-Renyi (ER) graphs with different number of nodes and average densities. There are seven real-world instances from the Zoo, all relatively small, with the biggest having only 754 nodes. Random networks are generated with 1000, 2000, 4000, 6000, 8000, and 10000  nodes and three different average density values. This dataset is further duplicated and adapted separately for the DRCR problem and for SRLG-disjoint DCR problem. For the latter, two different generators of SRLG are applied, following "random" and "star" patterns.

\begin{table}[!ht]
	\caption{Average number of links in each part of the generic dataset}
	\label{Table:Gen_dataset}
	\centering
	\begin{tabular}{|l|l|l|l|l|l|l|}
		\hline
		~ & \multicolumn{3}{c|}{Links, ER graphs} & \multicolumn{3}{c|}{Links, SF graphs} \\ \hline
		Nodes & k1 & k2 & k3 & m2 & m3 & m4  \\ \hline
		1000 & 6929 & 13833 & 20828 & 3992 & 5982 & 7968  \\ \hline
		2000 & 15293 & 29953 & 45625 & 7992 & 11982 & 15968  \\ \hline
		4000 & 33234 & 66240 & 99050 & 15992 & 23982 & 31968  \\ \hline
		6000 & 52470 & 104135 & 156321 & 23992 & 35982 & 47968  \\ \hline
		8000 & 72063 & 143249 & 215934 & 31992 & 47982 & 63968  \\ \hline
		10000 & 92110 & 184279 & 275652 & 39992 & 59982 & 79968  \\ \hline
		
	\end{tabular}
\end{table}

\begin{table}[!ht]
	\caption{Sizes of graphs from Topology Zoo}
	\label{Table:Zoo_dataset}
	\centering
	\begin{tabular}{|l|l|l|l|}
		\hline
		Name & Nodes & Links & Planar  \\ \hline
		Cogentco & 197 & 490 & Not  \\ \hline
		GtsCe & 149 & 386 & Yes  \\ \hline
		Interoute & 110 & 316 & Yes  \\ \hline
		Kdl & 754 & 1798 & Not  \\ \hline
		Pern & 127 & 258  & Yes \\ \hline
		TataNld & 145 & 388 & Yes  \\ \hline
		VtlWavenet2008 & 88 & 184 & Yes  \\ \hline
	\end{tabular}
\end{table}

"Star" SRLG pattern, also called "incident-SRLG", is widely known in literature on optical network survivability \cite{Xie:18}, \cite{Luo:05}, \cite{Datta:08}, \cite{Bermond:15}. For the "star" generator, links to include in an SRLG are chosen randomly from egress (outgoing) links of a particular node in the graph, and the size of an SRLG is randomly determined based on the average degree of the graph. For the "random" generator, the size of an SRLG is a random integer between 1 and 40, and links are randomly chosen across the whole graph to be included in the SRLG. The procedure is repeated until each link in the graph is included in at least one SRLG. Although we do not know literature except for the work \cite{Zhao:23} to explicitly study random SRLG, we believe that this rather simple technique may be an example of "non-star" pattern mentioned in work \cite{Datta:08}, and it provides a complementary example to the "star" pattern to improve diversity of instances in the dataset.

We should also mention another widely-known SRLG pattern, namely "regional" SRLGs \cite{Vass:22}, \cite{Berczi-Kovac:24}, \cite{Vass:20}, \cite{Vass:22-book}. Although being important for certain practical scenarios, it is mainly applicable to planar or nearly-planar network topologies and requires additional geographical information on the network node location. It is not present in dataset \cite{Drcr_Github} and is not used in our work.

The dataset \cite{Drcr_Github} contains 10 independent instances of random graphs for each combination of node number and density in DRCR case, making total of 180 graphs (10 random instances for each one of 6 node number options combined with 3 density options), and 3 independent instances for each set of parameters in SRLG case, making total of 108 graphs (3 random instances for each one of 6 node number options combined with 3 density options and 2 SRLG generators).
	
We extend the dataset, adding another important example of random networks, namely Barabasi-Albert graphs. Those graphs show scale-free behavior \cite{Barabasi:99}, therefore we call them scale-free (SF). Since the work of Barabasi and Albert, there has been a broad discussion on the extent to which such graphs reflect the properties of real-world networks. Not doing a complete review, we mention works \cite{Dorogovtsev:02}, \cite{Broido:19}, \cite{Holme:19} that give some insights into the discussion. We make a simple argument that this class of networks is important and serves as a good complement to the two classes of networks present in the original dataset. For a comparison of the three types of networks, see Appendix \ref{app:dataset}. We used the Python NetworkX library to generate SF networks \cite{networkx_ba} and followed the general design established in \cite{Zhao:23}. We also chose the number of nodes to be 1000, 2000, 4000, 6000, 8000, and 10000. For each number of nodes, we chose parameter m of the generator to be 2, 3, and 4  and generated 10 independent instances for each option. We turned the graphs to be directed by adding an edge in the opposite direction for each edge in the generated graph. Costs and delays are set to be independent uniformly distributed integers in the range $[1, ... 100]$. For SRLG problems, we chose the first five graphs in each category and independently generated random and star SRLGs. We generated 50 source-target pairs for each graph and assigned delay constraints for them. Finally, for the DRCR problem, we removed all tasks that appeared to have no feasible solution, and for SRLG-disjoint DCR problem, we kept only trap instances. After that, the total number of DRCR instances is 8498, the total number of trap instances for "random" SRLG is 3404, and for "star" SRLG is 3264. We consider non-trap instances trivial and thus not representative for evaluating the BTCS algorithm.

We show information on average number of links in each part of dataset in Tables \ref{Table:Gen_dataset}, \ref{Table:Zoo_dataset}. For the real-world networks we also include information of their planarity according to the paper \cite{Bowden11} as it can be important for further analysis.
	
Considering a realistic use case, where connections must be established promptly within hard time limits upon request, we propose as a primary metric for algorithm evaluation the number of feasible solutions found under 20 and 50 milliseconds because these values represent the real-life scenario. Our algorithms are particularly well-optimized to improve these metrics. We also provide more common measures, including maximal execution time, average execution time, and median time, for each type of network.
	
All experiments were performed on desktop computer with Intel(R) Core(TM) i7-10700 CPU and 32 Gb of memory, powered by Windows 10 Pro.

\subsection{DRCR}

In this section we evaluate and discuss the advantage of the BTBU technique for single path DRCR problem.
	
All instances of the DRCR problem in the dataset from \cite{Zhao:23} are relatively easy to solve for the Pulse+ algorithm. So, we do not split it into parts by topology types but generate statistics for the entire dataset altogether. This dataset has a total number of 11245 tasks, and the results are presented in Table \ref{Table:DRCR}.

\begin{table}[!ht]
	\caption{Performance comparison of DRCR algorithms on original dataset}
	\label{Table:DRCR}
	\centering
	\begin{tabular}{|l|l|l|l|l|}
		\hline
		~ & \makecell{Pulse+ \\ (Zhao \cite{Zhao:23})} & \makecell{Pulse+ \\ (replica)} & \makecell{BTBU \\ (option 1)} & \makecell{BTBU \\ (option 2)}  \\ \hline
		max time, ms & 87.25 & 42.00 & \textbf{22.49} & 24.70  \\ \hline
		average time, ms & 10.30 & 4.09 & \textbf{3.60} & 3.64  \\ \hline
		median time, ms & 6.32 & 1.89 & \textbf{1.65} & 1.70  \\ \hline
		solved in 20 ms & 9368 & 11089 & \textbf{11239} & 11237  \\ \hline
		solved in 50 ms & 11169 & 11245 & 11245 & 11245  \\ \hline
	\end{tabular}
\end{table}	 	

\begin{table}[!ht]
	\caption{Performance comparison of DRCR algorithms on SF networks}
	\label{Table:DRCR_SF}
	\centering
	\begin{tabular}{|l|l|l|l|l|}
		\hline
		~ & \makecell{Pulse+ \\ (Zhao \cite{Zhao:23})} & \makecell{Pulse+ \\ (replica)} & \makecell{BTBU \\ (option 1)} & \makecell{BTBU \\ (option 2)}  \\ \hline
		max time, ms & 2,030.21 & 686.43 & 206.50 & \textbf{172.44}  \\ \hline
		average time, ms & 12.00 & 3.54 & 2.19 & \textbf{1.91}  \\ \hline
		median time, ms & 5.28 & 2.08 & 1.68 & \textbf{1.54}  \\ \hline
		solved in 20 ms & 7562 & 8371 & 8483 & \textbf{8494}  \\ \hline
		solved in 50 ms & 8204 & 8462 & 8497 & 8497  \\ \hline
	\end{tabular}
	
\end{table}
	
We run four algorithms on this dataset. The first algorithm in the table is the Pulse+, implemented by the authors of \cite{Zhao:23} and obtained on Github \cite{Drcr_Github}. Second is the Pulse+, which we developed independently. Surprisingly, these two significantly differ in performance, which we assume to be due to implementation difference, since all main features of our Pulse+ implementation follow the ideas of \cite{Zhao:23}. Our code is built using Boost Graph Library, while the code in \cite{Drcr_Github} uses its own data structure for graph representation.
	 
To demonstrate the effect of artificial cost bounds, we compare our implementation of unbounded Pulse+ with the two techniques, described in section \ref{sec:bottom-top}. For the cost step value in the second technique we used the cost of the shortest edge in the graph multiplied by two.
	 
Including judiciously chosen bounds for cost improves the result, especially for the hardest instances. While average and median values are affected just a little; the maximal time is reduced almost by factor of 2. On the other hand, the whole dataset is not particularly difficult to solve. All 11245 instances are solved by all 3 of our algorithms within 50 milliseconds, no matter with or without artificial cost bounds.

In Table \ref{Table:DRCR_SF}, we present the results obtained on the SF dataset containing 8498 instances. We see that it is easier to solve on average, but the hardest instances take more time than in the dataset from \cite{Drcr_Github}. This can be expected because real-world networks in the first dataset are too small, and ER graphs have a uniform distribution of node degrees. At the same time, SF graphs have a power-law distribution of degrees, providing a wider range of task complexities. The overall pattern is similar to that in Table \ref{Table:DRCR}. Note that the second updating strategy for the cost bound (Algorithm \ref{alg:Pulse++2}) is more effective than the first one in this case.

\subsection{SRLG-disjoint DCR problem}

In this section we evaluate the performance of our BTCS algorithm (Algorithm \ref{alg:SRLG}) in comparison to COSE-Pulse+ for the SRLG-disjoint DCR problem.

\begin{table*}[!ht]
	
	\caption{Performance comparison of SRLG trap avoidance algorithms, real topologies instances with star SRLGs}
	\label{Table:SRLG_zoo_star}
	\centering
	\begin{tabular}{|l|l|l|l|l|}
		\hline
		~ & CoSE-Pulse+ & \makecell{BTCS-Pulse+ \\ (1 thread)}  & \makecell{BTCS-Pulse+ \\ (2 threads)} & \makecell{BTCS-Pulse+ \\ (4 threads)}  \\ \hline
		Instances, solved under 20 ms & 160 & 66 & 68 & 72  \\ \hline
		Instances, solved under 50 ms & 175 & 74 & 76 & 77  \\ \hline
		Total number of feasible solutions & 79 & 79 & 79 & 79  \\ \hline
		Feasible solutions found under 20 ms & 58 & 66 & 68 & 72  \\ \hline
		Feasible solutions found under 50 ms & 69 & 74 & 76 & 77 \\ \hline
		Max time to feasible solution, ms & 653.81 & 571.67 & 412.18 & 223.19  \\ \hline
		Average time to feasible solution, ms & 29.35 & 17.69 & 13.07 & 8.78  \\ \hline
		Median time to feasible solution, ms & 6.3234 & 3.2595 & 2.3602 & 1.9091  \\ \hline
	\end{tabular}
	
\end{table*}

\begin{table*}[!ht]
	\caption{Performance comparison of SRLG trap avoidance algorithms, ER instances with random SRLGs}
	\label{Table:SRLG_gen_rand}
	\centering
	\begin{tabular}{|l|l|l|l|l|}
		\hline
		~ & CoSE-Pulse+ & \makecell{BTCS-Pulse+ \\ (1 thread)}  & \makecell{BTCS-Pulse+ \\ (2 threads)} & \makecell{BTCS-Pulse+ \\ (4 threads)}  \\ \hline
		Instances, solved under 20 ms & 73 & 71 & 79 & 87  \\ \hline
		Instances, solved under 50 ms & 107 & 101 & 106 & 112  \\ \hline
		Total number of feasible solutions & 140 & 140 & 140 & 140  \\ \hline
		Max time, ms & 171.995 & 226.466 & 230.338 & 168.566  \\ \hline
		Average time, ms & 36.14 & 37.43 & 33.20 & 28.16  \\ \hline
		Median time, ms & 18.51 & 19.63 & 14.88 & 13.01  \\ \hline
	\end{tabular}
	
	\bigskip
	
	\caption{Performance comparison of SRLG trap avoidance algorithms, ER instances with star SRLGs}
	\label{Table:SRLG_gen_star}
	\centering
	\begin{tabular}{|l|l|l|l|l|}
		\hline
		~ & CoSE-Pulse+ & \makecell{BTCS-Pulse+ \\ (1 thread)}  & \makecell{BTCS-Pulse+ \\ (2 threads)} & \makecell{BTCS-Pulse+ \\ (4 threads)}  \\ \hline
		Instances, solved under 20 ms & 77 & 96 & 99 & 101  \\ \hline
		Instances, solved under 50 ms & 133 & 149 & 153 & 158  \\ \hline
		Total number of feasible solutions & 212 & 212 & 212 & 212  \\ \hline
		Max time, ms & 1310.06 & 2337.2 & 1597.66 & 1352.32  \\ \hline
		Average time, ms & 68.18 & 62.65 & 52.98 & 45.24  \\ \hline
		Median time, ms & 30.94 & 24.21 & 21.53 & 21.21  \\ \hline
	\end{tabular}
	
\end{table*}

\begin{table*}[!ht]
	
	\caption{Performance comparison of SRLG trap avoidance algorithms, SF instances with random SRLGs}
	\label{Table:SRLG_SF_rand}
	\centering
	\begin{tabular}{|l|l|l|l|l|}
		\hline
		~ & CoSE-Pulse+ & \makecell{BTCS-Pulse+ \\ (1 thread)}  & \makecell{BTCS-Pulse+ \\ (2 threads)} & \makecell{BTCS-Pulse+ \\ (4 threads)}  \\ \hline
		Instances, solved under 20 ms & 1645 & 2340 & 2524 & 2420  \\ \hline
		Instances, solved under 50 ms & 2811 & 3037 & 3042 & 3013  \\ \hline
		Total number of feasible solutions & 3210 & 3210 & 3210 & 3210  \\ \hline
		Feasible solutions found under 20 ms & 1509 & 2340 & 2524 & 2420  \\ \hline
		Feasible solutions found under 50 ms & 2634 & 3037 & 3042 & 3013  \\ \hline
		Max time to feasible solution, ms & 30840 & 25843.2 & 24912 & 24568.2  \\ \hline
		Average time to feasible solution, ms & 138.75 & 54.13 & 49.93 & 53.40  \\ \hline
		Median time to feasible solution, ms & 17.69 & 10.89 & 9.88 & 10.46  \\ \hline
	\end{tabular}
	
	\bigskip
	
	\caption{Performance comparison of SRLG trap avoidance algorithms, SF instances with star SRLGs}
	\label{Table:SRLG_SF_star}
	\centering
	\begin{tabular}{|l|l|l|l|l|}
		\hline
		~ & CoSE-Pulse+ & \makecell{BTCS-Pulse+ \\ (1 thread)}  & \makecell{BTCS-Pulse+ \\ (2 threads)} & \makecell{BTCS-Pulse+ \\ (4 threads)}  \\ \hline
		Instances, solved under 20 ms & 1571 & 2541 & 2733 & 2620  \\ \hline
		Instances, solved under 50 ms & 2705 & 2959 & 2963 & 2946  \\ \hline
		Total number of feasible solutions & 3114 & 3114 & 3114 & 3114  \\ \hline
		Feasible solutions found under 20 ms & 1527 & 2541 & 2733 & 2620  \\ \hline
		Feasible solutions found under 50 ms & 2638 & 2959 & 2963 & 2946  \\ \hline
		Max time to feasible solution, ms & 2241.75 & 11942.8 & 14107 & 18253.6  \\ \hline
		Average time to feasible solution, ms & 29.01 & 34.66 & 35.81 & 43.41  \\ \hline
		Median time to feasible solution, ms & 19.10 & 10.19 & 8.27 & 9.08  \\ \hline
	\end{tabular}
	
\end{table*}
	
To highlight algorithm performance when dealing with different network topologies and different SRLG assignment options, we present the results separately for real-world, ER, and SF networks, each with random and star SRLG generators. There are six combinations, and the results are listed in tables \ref{Table:SRLG_zoo_star} - \ref{Table:SRLG_SF_star}. The only combination missing in these tables is real-world networks with random SRLG, which has no feasible solutions. To show the advantage of parallel execution of BTCS and give further insight into its nature, we run it in 1, 2 and 4 parallel threads. For the corridor width in all experiments we use the cost of the shortest edge in the graph, multiplied by 10.
	
Table \ref{Table:SRLG_zoo_star} contains results for Zoo instances with "star" SRLG. There are 210 trapped instances, 79 of which are avoidable traps. The result is most representative of the features of the BTCS algorithm, as it was expected to be. It has a significant advantage in the search for feasible solutions, while CoSE-Pulse+ is much better when dealing with unavoidable traps and can provide proof of infeasibility very fast. It manifests in a higher total number of solved instances for CoSE-Pulse+, including feasible and infeasible ones. At the same time, BTCS finds more feasible solutions and has a lower average time of the search for feasible paths. Parallel execution improves the time of search and the number of feasible solutions found under hard time limits.

Tables \ref{Table:SRLG_gen_rand} and \ref{Table:SRLG_gen_star} present results for ER instances with random and star SRLG assignment. There are a total of 140 and 212 trapped instances, correspondingly. All those instances have feasible solutions. The results here are not as straightforward as those of the real-world topologies. For random SRLG, BTCS can beat CoSE-Pulse+ only in parallel implementation. For star SRLG, BTCS outperforms CoSE-Pulse+ regarding the number of feasible solutions found in 50 and 20 ms but is inferior in execution time for the most complex instance - even in multithread implementation.

Results for SRLG-disjoint DCR problem on SF networks are presented in Tables \ref{Table:SRLG_SF_rand}, \ref{Table:SRLG_SF_star}. This part of the dataset has 3404 and 3264 trapped problems for random and star SRLG, respectively. Most of them are avoidable traps. The BTCS algorithm, even in one thread, clearly outperforms CoSE-Pulse+ in terms of the number of instances solved in 20 and 50 milliseconds. At the same time, in the case of star SRLG, BTCS is inferior in terms of the maximal and average time of the search for a feasible solution. Most interestingly, increasing the number of threads does not always improve algorithm performance. We will discuss it further in the next section.

\subsection{Analysis and discussion}
	\label{sec:discussion}

In the previous section, we presented and briefly described the experimental data on the performance of the BTCS algorithm and compared it to the CoSE-Pulse+ algorithm. In this section, we will discuss the results from three different points of view: single-thread performance, features of multithread execution, and worst-case performance.

\subsubsection{Single-thread performance}
	
All results in the previous section are presented for single- and multithread implementation of the BTCS algorithm. Importantly, considering the number of feasible solutions found under the limit of 20 or 50 ms, BTCS algorithm in most cases has advantage over CoSE-Pulse+ even in single-thread version. The only exception is ER instances with random SRLG, where the single-thread performance is inferior (though still comparable). Considering other metrics, such as maximal, average and median time to feasible solution, BTCS in single thread also has comparable performance to CoSE-Pulse+, and outperforms it in certain scenarios (see Tables \ref{Table:SRLG_zoo_star} - \ref{Table:SRLG_SF_star} for details). This observation underpins the claim that the competitiveness of BTCS originates from its basic principles and not from any sophisticated multithread implementation. The only exception is the SF instances with star SRLG, where the maximal execution time of BTCS in single thread is almost 6x worse than that of CoSE-Pulse+.

\subsubsection{Multithread performance}

The multithread execution results in the previous section are obtained with a simple and relatively straightforward approach. In this implementation, the algorithm performs parallel search for AP candidates in several subsequent cost corridors. Other variants, such as the parallel search for disjoint PP within a corridor, are possible, as mentioned in Section \ref{sec:corridor}. Conflicting link paradigm also allows parallelization, as in \cite{Xie:18}, although it was not generalized to handle delay-constrained problems to the best of our knowledge. We consider detailed research of parallel execution capabilities for the SRLG-disjoint DRCR problem a subject of future work.
	
The approach implemented in the present work allows further insight into how the distribution of feasible solutions influences the algorithm's performance. Note that Tables \ref{Table:SRLG_zoo_star} - \ref{Table:SRLG_gen_star} show "normal" behavior, i. e. the number of found feasible solutions increases and execution time decreases with growing number of threads. On the contrary, Tables \ref{Table:SRLG_SF_rand} - \ref{Table:SRLG_SF_star} show "abnormal" behavior, namely, the performance deteriorates between two and four threads. We attribute this issue to the path distribution relative to cost. Too many threads means that the higher thread will often search in the space above the shortest feasible solution. Since there are more AP candidates at higher cost values, these threads will take more time, slowing down the whole process.
	
\begin{figure*}[!t]
	\centering
	\includegraphics[width= 1 \textwidth]{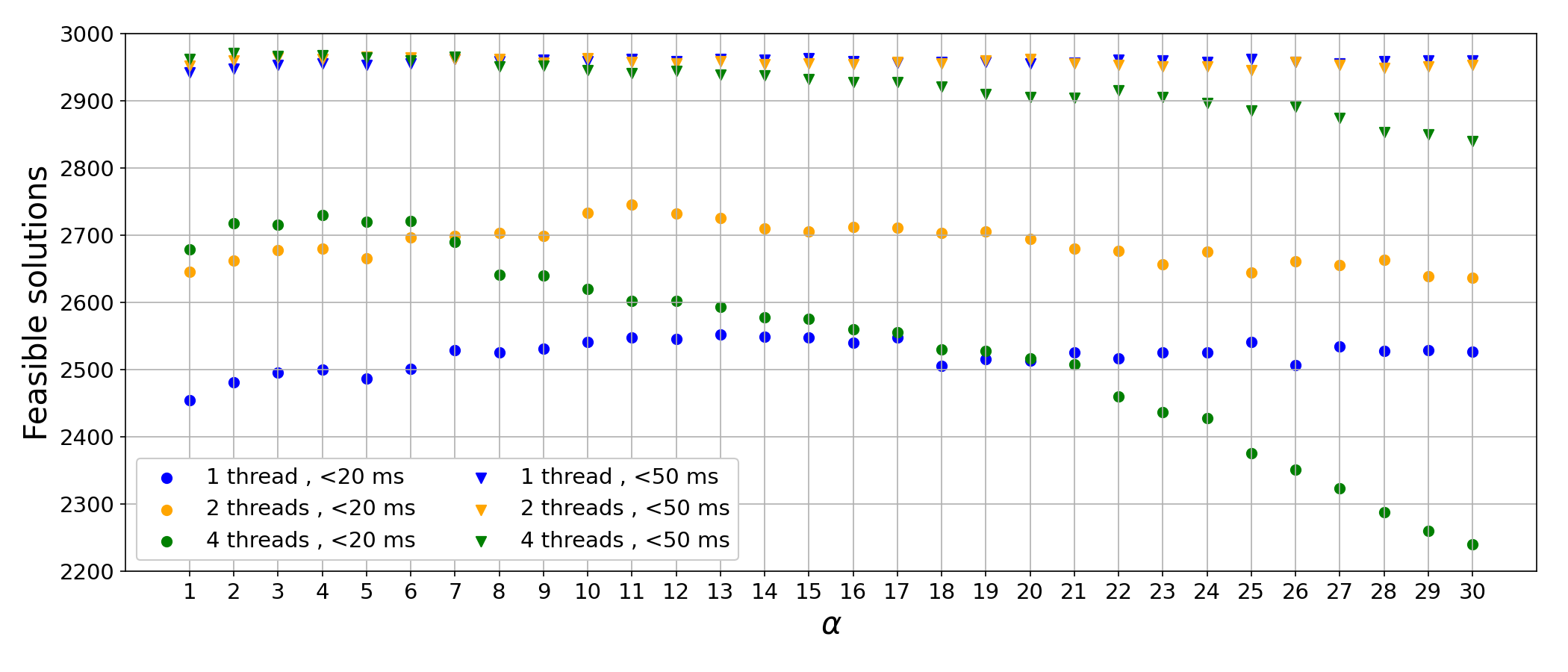}
	\caption{Number of feasible paths, found by BTCS algorithm, with respect to the corridor width parameter $\alpha$. The dataset used is SF graphs with "star" SRLG generator. See the results for other graph topologies and SRLG distributions in Appendix \ref{app:alpha_dynamics}.}
	\label{fig:alpha_dynamics}
\end{figure*}
	
To understand this phenomenon more deeply, we introduce parameter $\alpha$ to account for the corridor width in the algorithm implementation. The width of the corridor is the cost of the shortest edge in the graph, multiplied by $\alpha$.	So, the experiments performed in the previous section correspond to $\alpha=10$. In Figure \ref{fig:alpha_dynamics}, we plot the number of feasible paths found under 20 and 50 milliseconds as a function of $\alpha$, which is effectively the corridor width. The figure shows that algorithm performance in one and two threads is quite stable with respect to $\alpha$ and changes only a little in the wide range of parameter values. The value $\alpha = 10$ is close to optimal for these two options, and double-thread implementation has clear advantage over single-thread. At the same time, the four-thread implementation has no overall advantage compared to double-thread, and its performance decays quickly with growing $\alpha$. 

Similar analysis for other types of networks in the dataset can be found in Appendix \ref{app:alpha_dynamics}. Since in real applications the network topology can be known in advance, we suggest running experiments and selecting algorithm parameters that are best for actual network.

\subsubsection{Worst-case performance}

Although worst-case performance is interesting theoretically, it is a less valuable metric in real applications. For an algorithm intended to be applied in the operation of modern networks, with acceptable delays measured in milliseconds, an algorithm that solves a problem in a second does not gain much advantage over the one that takes 5 seconds for the same task. This is why we decided to primarily compare algorithms by the number of problems solved in 20 and 50 milliseconds. Nonetheless, we provide and discuss some data below to give insight into the algorithm's worst-case performance.
	
\begin{figure*}[!ht]
	\centering
	\includegraphics[width= 1 \textwidth]{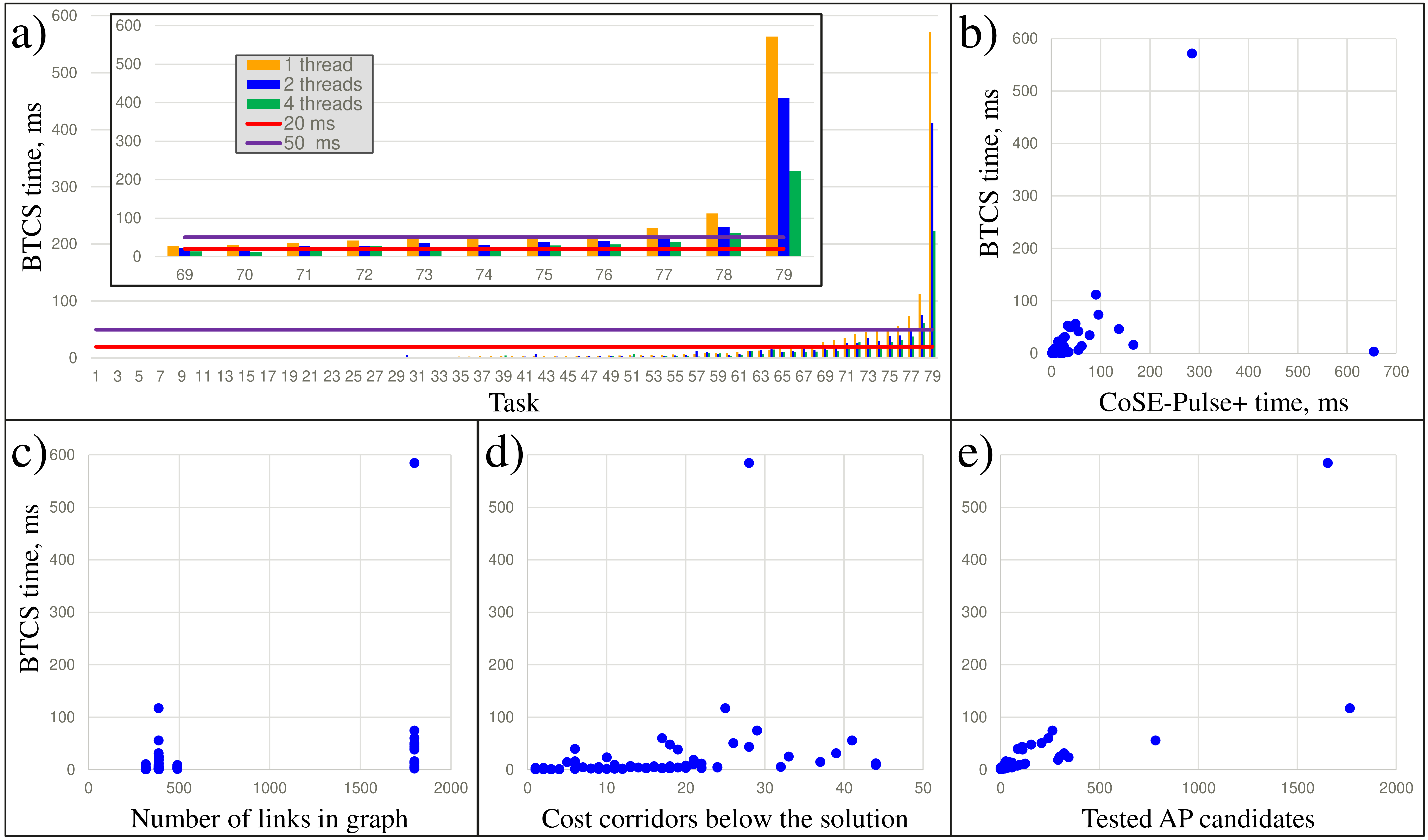}
	\caption{Distribution of solution times for SRLG-disjoint DCR problem on the real-world network instances. (a): absolute execution time of BTCS algorithm on Zoo dataset with star SRLG. (b): distribution of execution times for CoSE-Pulse+ algorithm against single-thread BTCS. (c): BTCS algorithm execution time dependency on the graph size. (d): BTCS algorithm execution time dependency on the number of cost corridors explored before the solution is found. (e): BTCS algorithm execution time dependency on the number of AP candidates that are shorter than the solution, but have no disjoint protection.}
	\label{fig:SRLG_tasks}
\end{figure*}

Figure \ref{fig:SRLG_tasks} (a) shows all tasks, corresponding to the Table \ref{Table:SRLG_zoo_star}, sorted by their execution times in ascending order. It demonstrates that majority of tasks in the dataset are easy, taking only a few milliseconds to find a solution. Some fraction of tasks (shown on the inset) are considerably hard, with solution times measured in tens of milliseconds, and only a few of them make the worst case, with solution times reaching hundreds of milliseconds. Figure \ref{fig:SRLG_tasks} (b) compares solution times of CoSE-Pulse+ and BTCS for all these tasks. It shows that in general tasks that are hard for CoSE, also tend to be hard for BTCS. On the other hand, execution time for a particular task may significantly differ for the two algorithms.

Figure \ref{fig:SRLG_tasks} (c) - (e) dives into possible explanations for the difficulty variation of these tasks (we provide similar data for other graph topologies in Appendix \ref{app:worst_case}). As a general conclusion, no single parameter would explain the worst-case behavior of the BTCS algorithm. The number of cost corridors to be explored by the algorithm is limited to a few dozen for all problem instances. There is a strong correlation between execution time and the number of AP candidates that has to be considered before the solution is found, but the correlation strength is different for each particular graph. Also, more complex tasks tend to appear on the bigger graphs, which may explain the correlation in difficulty between BTCS and CoSE. Other metrics, such as graph density, centrality, or degrees of source and target nodes, may also contribute to the difficulty of a particular task.
	
\begin{figure}[!ht]
	\centering
	\includegraphics[width=0.47 \textwidth]{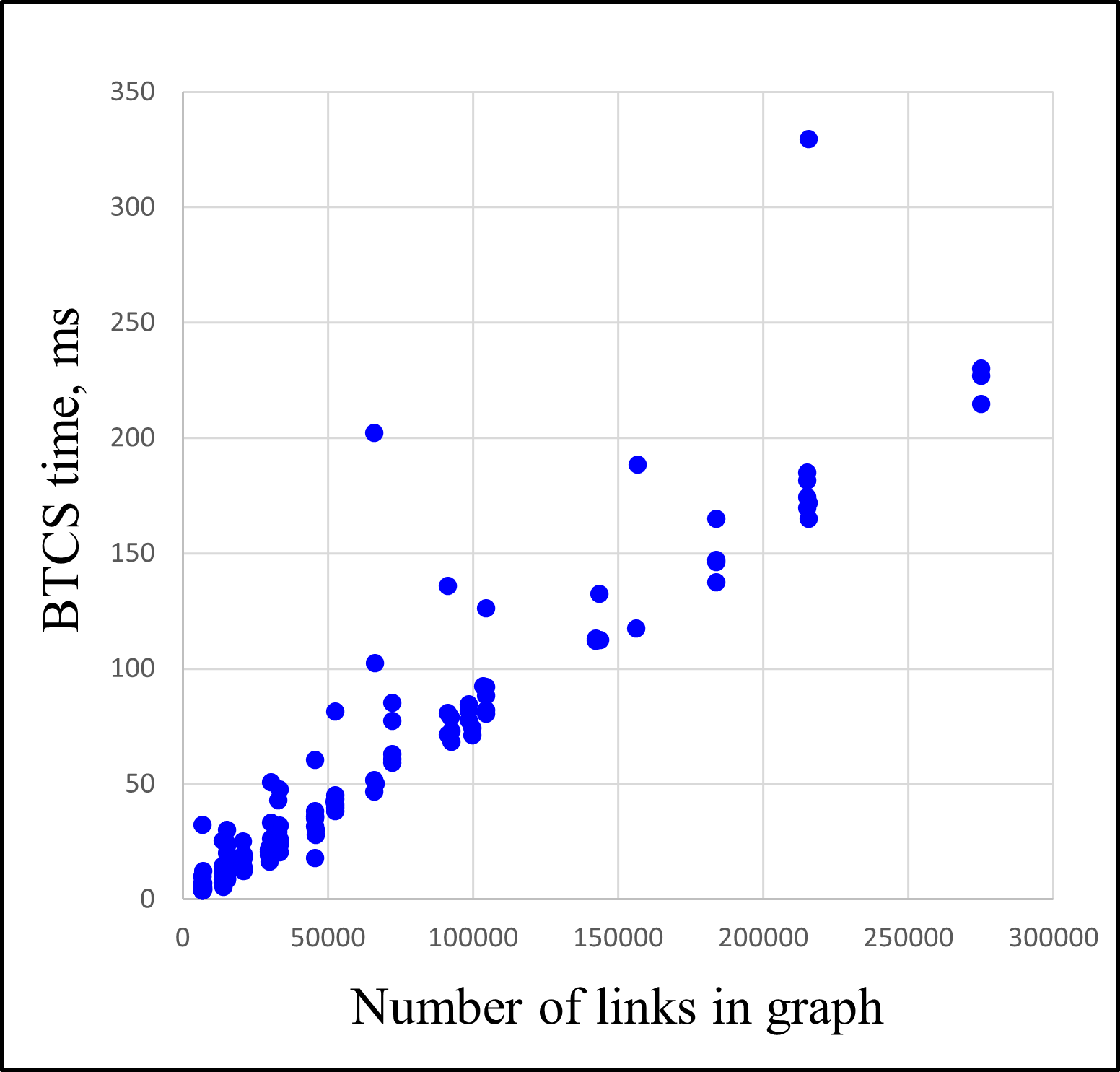}
	\caption{BTCS algorithm execution time dependency on the graph size for ER random graphs with random SRLG.}
	\label{fig:ER_random_link_num_corr}
\end{figure}

Figure \ref{fig:ER_random_link_num_corr} shows correlation between number of links in a graph to the execution time of BTCS algorithm on ER topologies with random SRLG. There is similar though not so explicit tendency on other network and SRLG types. More information can be found in Appendix \ref{app:worst_case}.
	
This uncertain nature of the problem hardness may be explained by the fact that the algorithm has two levels of complexity. At the higher level, BTCS explores the search space in cost corridors, while at the lower level, each particular path is found by variants of the Pulse+ algorithm, which relies on more subtle features of the graph rather than statistical properties of path distribution.
	
\section{Conclusion and future work}
	
	In this paper, we propose two algorithms for two routing problems necessary to solve when operating modern Internet networks. Our approach is based on statistical analysis of the typical path distribution and takes advantage of additional pruning by cost for the classical Pulse / Pulse+ algorithm. Both algorithms are competitive with previously known ones, which were verified in rigorous experiments. Particularly, our algorithms are the best choice when the optimal solution is required in a short period of time upon request. Our BTCS algorithm for trap avoidance for SRLG-disjoint DRCR problem is innovative, as it presents an approach tangential to the previously dominant paradigm of conflicting set exclusion algorithms. It is based on performing searches in corridors, which are a function of cost, and exploiting additional pruning strategy to reduce search space instead of exploiting entire sets of conflicting SRLGs. 
	
	We suggest several ways of future improvement. One of them is application of the technique inspired by the queuing procedure from paper \cite{Bolivar:14}. The original procedure stops pulses that have reached certain number of links, placing them in a queue and reusing again after all pulses have reached the limit. This approach turns the DFS search in Pulse algorithm into a combination of DFS and BFS, making it pure BFS when the depth limit is set to 1. This can be used as an inspiration in the search for AP candidates in BTCS, by placing the pulses that have reached the upper bound of current cost corridor in a queue and reusing in the next corridor instead of pruning them. It will improve speed at the price of memory to store all the pulses, that otherwise would be pruned.
	
	Also it may be possible to use observations on search space structure made in this work to speed up the CoSE algorithm, or algorithms, developed for planar networks in \cite{Vass:22} and \cite{Berczi-Kovac:24}. A merging of these techniques, exploiting conflicting link sets or leveraging planar topology together with judiciously chosen cost bounds for efficient pruning, may result in further progress in the area.
	
	Another way of improvement, considering the multithread implementation of the algorithm, is the development of more sophisticated ways of interaction between threads that will allow to stop execution when the solution is found in any thread. It will also allow mitigation of the performance decline for higher number of threads, observed on SF networks. Carefully designing breaking rules for multithread execution, one can also turn the algorithm into an approximate one, trading exactness of the solution for speed. For example, it can be achieved by independent exploration of odd and even corridors in two parallel threads, allowing one thread not to wait for the other on each step and breaking at the time when the first protected pair is found. Since the exploration goes in the bottom-top direction, the obtained solution will be relatively close to the optimal. We also can think of including parallelism on different level, and instead of searching for candidates in parallel corridors, search for protection paths in parallel for all AP candidates in each corridor, as was discussed in Section \ref{sec:bottom-top}.
	
	Finally, in the process of this work we bumped into the question of having a reliable standard and realistic dataset to evaluate algorithms for modern communication standards. As we showed in the Experiment section, all available real-world networks are relatively small and easy to solve. The Topology Zoo collection was first published in 2011, and hardly reflects modern state of the networking technology. It also lacks of important data, such as real costs, delays and SRLG distributions. The Topology Zoo has been analyzed in the work \cite{Bowden11} and the majority of networks were found to be planar. The same work observed the tendency that bigger networks are more often non-planar. The question remains, is it actually the case with real networks containing thousands of nodes?
	
	In our work we generated costs, delays and SRLG, following the rules proposed in \cite{Zhao:23}. Although we tried our best to test different network topologies and SRLG patterns, it is still possible that the path distribution with cost on more realistic networks will not benefit our algorithm to the same degree as the generic dataset used in this work. Nonetheless, we still believe that any pruning-based algorithm can make use of the analysis of such distribution, as we did in this work for the first time.

\section*{Acknowledgments}
We thank our colleague Yuriy Zotov for his essential help with C++ code implementation and optimization at the early stages of the work.

\clearpage

\onecolumn

\begin{appendices}
	
	\section{Pulse+ algorithm and its modifications}
	\label{app:pulse+}
	
	In this section we describe and provide pseudocode for 3 modifications of Pulse+ algorithm, used as a subroutine in the BTBU and BTCS algorithms. 
	
	\subsection{Pulse+ algorithm}
	\label{app:pulse+orig}
	
	The pseudocode, presented here, corresponds to the original Pulse+ algorithm, proposed in paper \cite{Zhao:23}. It is intended to solve Delay-Range Constrained Routing problem and return the optimal path for it. In our paper it is used twice: in section \ref{sec:bottom-top} (Algorithms \ref{alg:Pulse++1} and \ref{alg:Pulse++2}) and in Algorithm \ref{alg:SRLG}. We divide the code into two parts for convenience: the high-level code (Algorithm \ref{alg:Pulse+_high}) and details of recursive Pulse function (Algorithm \ref{alg:Pulse+_recursive}).
	
	\begin{algorithm}
		
		\SetKwInOut{Input}{input}\SetKwInOut{Output}{output}
		\SetAlgoLined
		\Input{Graph G, source $s$, target $t$, delay range $[d_{low}, d_{up}]$.}
		\Output{The shortest path from $s$ to $t$, satisfying delay range constraint}
		
		$current\_min\_cost = init\_value$  /* \textit{infinity for original Pulse+ or given by upper cost bound in our techniques} */
		
		$P = \{\}$
		
		$P^* = \{\}$
		
		$visited\_nodes = \{\}$
		
		$current\_cost = 0$
		
		$current\_delay = 0$
		
		Initialization(G, $t$).  /* \textit{calculate min cost and min delay trees from each node to target with the help of Dijkstra or A* algorithm} */
		
		Pulse\_plus($P^*$, $P$, $s$, $visited\_nodes$, $current\_cost$, $current\_delay$, $current\_min\_cost$)
		
		\Return current\_min\_cost, $P^*$
		
		\caption{Pulse+ algorithm}
		\label{alg:Pulse+_high_app}
	\end{algorithm}

	\begin{algorithm}
		
		\SetKwInOut{Input}{input}\SetKwInOut{Output}{output}
		\SetAlgoLined
		\Input{current optimal path $P^*$, current path $P$, current node $n$, list of visited nodes $visited\_nodes$, current cost $c$, current delay $d$, $current\_min\_cost$}
		\Output{void}
		
		\If{$n == t$ \textbf{and} $d_{low} \le d \le d_{up}$ \textbf{and} $c < current\_min\_cost$}{
			$P^* = P$
			
			$current\_min\_cost = c$
			
			\Return
		}
		
		\For{$(nk)$ in egress links of node $n$}{
			\lIf{$k$ \textbf{in} $visited\_nodes$}{continue}
			\lIf{$d + d(nk) + d(P^{min\_delay}_{k \rightarrow t}) > d_{up}$}{\label{alg:pulse+:infeas_app}continue}
			\lIf{$c + c(nk) + c(P^{min\_cost}_{k \rightarrow t}) > current\_min\_cost$}{\label{alg:pulse+:opt_app}continue}

			Pulse\_plus($P^*$, $P + (nk)$, $k$, $visited\_nodes + \{k\}$, $current\_cost + c(nk)$, $current\_delay + d(nk)$, $current\_min\_cost$)
			
		}
		\caption{Pulse\_plus()  recursive function}
		\label{alg:Pulse+_recursive_app}
	\end{algorithm}
	
	In the initialization phase of algorithm \ref{alg:Pulse+_high}, the minimal cost and minimal delay paths from each node to the target node are computed, similar to original Pulse algorithm \cite{Lozano:13}.
	
	Lines \ref{alg:pulse+:infeas} and \ref{alg:pulse+:opt} of algorithm \ref{alg:Pulse+_recursive} contain infeasibility and optimality prunings, respectively. Dominance pruning is not applicable to the scenario with nonzero $d_{low}$, so, generally this algorithm is not as efficient as original Pulse. In order to avoid loops in found paths, it is also essential to track visited nodes in this scenario.
	
	\subsection{Pulse+ modification for all paths search in cost corridor}
	\label{app:pulse+all}
	
	On the line \ref{algSRLG:line:all} of algorithm \ref{alg:SRLG} we use modification of Pulse+ algorithm to save all paths in current cost corridor. We put it into Appendix, because the difference with original Pulse+ is minor. 
	
	The algorithm explores the search space in the same manner with Pulse+, and uses same pruning strategies. Only now cost and delay play absolutely symmetric role. For both of them we have lower and upper bound, which is provided in problem statement for delay and artificially introduced on the higher level of the computation for cost. Upper limits for both delay and cost are used in prunings, while lower limits are only used in path selection when a pulse reaches the target node. Compared to the original Pulse+, we do not update upper limit on cost and the best found path. Instead, we keep cost limits constant and save all selected paths as future AP candidates.
	
	\begin{algorithm}[!ht]
		
		\SetKwInOut{Input}{input}\SetKwInOut{Output}{output}
		\SetAlgoLined
		\Input{Graph G, source $s$, target $t$, delay range $[d_{low}, d_{up}]$, cost corridor $[c_{low}, c_{up}]$.}
		\Output{All paths from $s$ to $t$ within delay range and cost corridor}
		
		$P = \{\}$
		
		$Paths = []$
		
		$visited\_nodes = \{\}$
		
		$current\_cost = 0$
		
		$current\_delay = 0$
		
		Initialization(G, $t$).
		
		Pulse\_plus\_all($Paths$, $P$, $s$, $visited\_nodes$, $current\_cost$, $current\_delay$)
		
		\Return $Paths$
		
		\caption{Pulse+ algorithm for all paths in cost corridor}
		\label{alg:Pulse+_all}
	\end{algorithm}
	
	\begin{algorithm}[!ht]
		\label{alg:pulse+_all:infeas}
		\SetKwInOut{Input}{input}\SetKwInOut{Output}{output}
		\SetAlgoLined
		\Input{set of AP candidates found so far $Paths$, current path $P$, current node $n$, list of visited nodes $visited\_nodes$, current cost $c$, current delay $d$}
		\Output{void}
		
		\If{$n == t$ \textbf{and} $d_{low} \le d \le d_{up}$ \textbf{and} $c_{low} \le c < c_{up}$}{
			$Paths.add(P)$
			
			\Return
		}
		
		\For{$(nk)$ in egress links of node $n$}{
			\lIf{$k$ \textbf{in} $visited\_nodes$}{continue}
			\lIf{$d + d(nk) + d(P^{min\_delay}_{k \rightarrow t}) > d_{up}$}{continue}
			\lIf{$c + c(nk) + c(P^{min\_cost}_{k \rightarrow t}) > c_{up}$}{\label{alg:pulse+_all:opt}continue}	
			
			Pulse\_plus\_all($Paths$, $P + (nk)$, $k$, $visited\_nodes + \{k\}$, $current\_cost + c(nk)$, $current\_delay + d(nk)$)
			
		}
		\caption{Pulse\_plus\_all()  recursive function}
		\label{alg:Pulse+_recursive_all}
	\end{algorithm}
	
	\newpage

	\subsection{Pulse+ algorithm for first feasible path}
	\label{app:pulse+feas}
	
	In this section we describe, perhaps, the simplest implementation of the Pulse+ algorithm, used in algorithm \ref{alg:SRLG} on line \ref{algSRLG:line:PP_search}. It is intended to return any feasible path as fast as possible. So, it does not keep track of cost value, and applies only infeasibility pruning. It also must terminate the recursion, when the first feasible path was found. The function on line \ref{algfeas:terminate} of algorithm \ref{alg:Pulse+_recursive_feas} is responsible for it, but we do not go into details of implementation. On practice it can be done by inclusion of a global boolean flag or by any other mean.

	\begin{algorithm}
		
		\SetKwInOut{Input}{input}\SetKwInOut{Output}{output}
		\SetAlgoLined
		\Input{Graph G, source $s$, target $t$, delay range $[d_{low}, d_{up}]$.}
		\Output{A path from $s$ to $t$, satisfying delay range constraint}
		
		$P = \{\}$
		
		$P^* = \{\}$
		
		$visited\_nodes = \{\}$
		
		$current\_delay = 0$
		
		Initialization(G, $t$).
		
		Pulse\_plus($P^*$, $P$, $s$, $visited\_nodes$, $current\_delay$)
		
		\Return $P^*$
		
		\caption{Pulse+ algorithm}
		\label{alg:Pulse+_feas}
	\end{algorithm}

	\begin{algorithm}
		
		\SetKwInOut{Input}{input}\SetKwInOut{Output}{output}
		\SetAlgoLined
		\Input{$P^*$, current path $P$, current node $n$, list of visited nodes $visited\_nodes$, current delay $d$}
		\Output{void}
		
		\If{$n == t$ \textbf{and} $d_{low} \le d \le d_{up}$}{
			$P^* = P$
			\textbf{Terminate recursion}
			\label{algfeas:terminate}
		}
		
		\For{$(nk)$ in egress links of node $n$}{
			\lIf{$k$ \textbf{in} $visited\_nodes$}{continue}
			\lIf{$d + d(nk) + d(P^{min\_delay}_{k \rightarrow t}) > d_{up}$}{continue}
			
			Pulse\_plus($P^*$, $P + (nk)$, $k$, $visited\_nodes + \{k\}$, $current\_delay + d(nk)$)
			
		}
		\caption{Pulse\_plus()  recursive function}
		\label{alg:Pulse+_recursive_feas}
	\end{algorithm}

	\newpage
	
	\section{Features of dataset}
	\label{app:dataset}
	
	In order to justify inclusion of SF graphs into the dataset, we draw the node degree distribution for typical ER graph, SF graph and realistic topology graph from Topology Zoo. In order to provide reasonable comparison, we have chosen graphs with almost similar number of nodes. Namely, "Kdl" graph from Topology Zoo, having 754 nodes (it is the biggest one in the Zoo), and two instances with 1000 nodes from ER and SF datasets.

\begin{figure*}[!ht]
		\centering
		\includegraphics[width= 1 \textwidth]{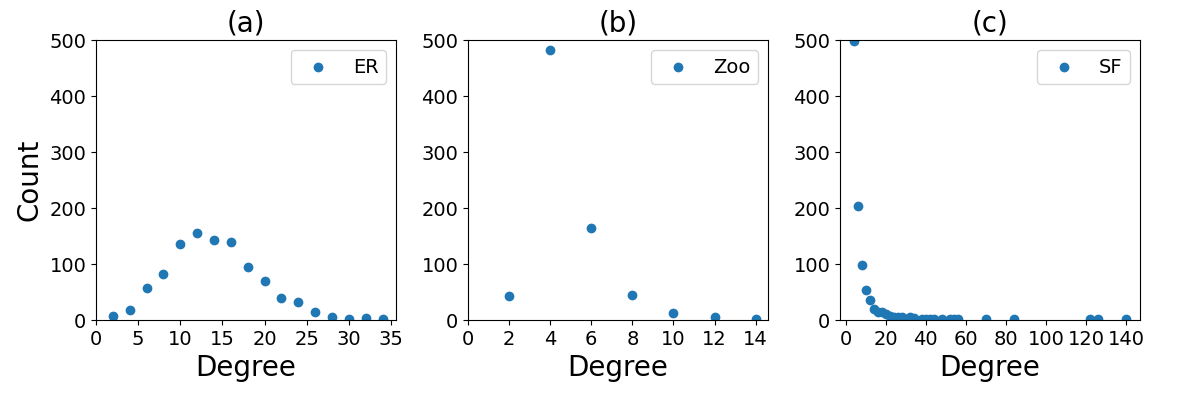}
		\caption{Distribution of node degrees for typical graphs from the 3 parts of the dataset. (a): ER graph, having 1000 nodes. (b): Kdl graph from Topology Zoo, having 754 nodes. (c): SF graph, having 1000 nodes.}
		\label{fig:graph_degree}
	\end{figure*}

	From Figure \ref{fig:graph_degree} we clearly see the distinction between three graph types. ER graph has wide degree distribution with no peaks and most nodes having degrees between 10 and 20. Kdl graph is sparse, with most nodes having degree 4 and highest degree 14. SF graph has long tail distribution, with few "hub" nodes having extremely high degrees up to 140, while majority of nodes have degree 2.
	
	This picture clearly shows, that SF graphs represent a class of networks, sufficiently different from both ER and Zoo graphs.
	
	\clearpage
	
	\section{Performance of BTCS algorithm with respect to the corridor width}
	\label{app:alpha_dynamics}
	
\begin{figure*}[!ht]
		\centering
		\includegraphics[width= 1 \textwidth]{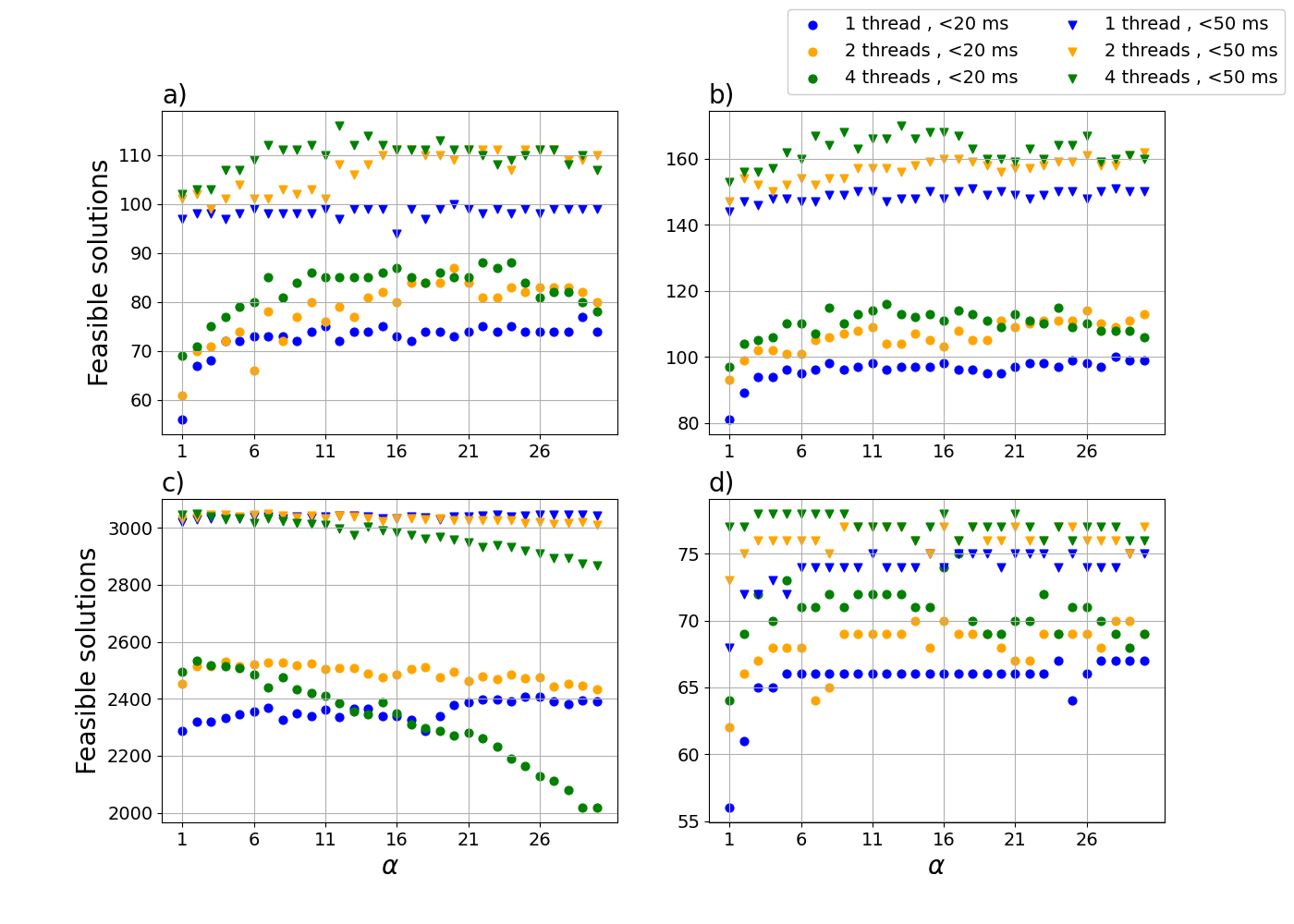}
		\caption{Number of feasible paths, found by BTCS algorithm, with respect to the corridor width parameter $\alpha$. a) Erdos-Renui graphs with random SRLG generator, b)Erdos-Renui graphs with star SRLG generator, c) SF graphs with random SRLG generator, d) Real-world graphs from Topology Zoo with star SRLG generator}
		\label{fig:alpha_dynamics_sup_combined}
	\end{figure*}

	In this section we provide more data on BTCS algorithm performance with respect to corridor width. We use same implementation with the section \ref{sec:discussion}, where the width of the corridor equals to the cost of the shortest edge in the graph, multiplied by $\alpha$. Figure \ref{fig:alpha_dynamics_sup_combined} a) - d) covers all other dataset parts, that are not presented in section \ref{sec:discussion}. This data indicates, that for all graph topologies, except SF, multiple thread execution is beneficial, and algorithm performance with respect to $\alpha$ is quite stable. The only strong recommendation is, that $\alpha$ should not be too small. At the same time, for SF graph topology with random SRLGs, the result is similar to that discussed in section \ref{sec:discussion}. For this type of graphs, double-thread implementation has clear advantage over single-thread, while four-thread implementation does not bring further improvement and quickly loses performance with growing $\alpha$.
	
	\clearpage
	
	\section{Low-cost tails of the path distributions for CSP and SRLG-disjoint problems}
	\label{app:tails}
	
	In this section we show low-cost tails of path distributions for both DRCR and SRLG-disjoint DCR problems, that are generated using networks of different topologies and sizes. This must support our claim, that distributions on figures \ref{fig:CSP_distribution} and \ref{fig:SRLG_space} are representative and show, that the motivation behind our algorithms for both problems makes sense for many different networks.
	
	We decided to plot only low cost ends of the distributions, since both our algorithms use artificial bounds on cost and thus effectively work only with paths of the lower costs. We use logarithmic scale on the Y-axis of all plots in order to make diagrams easier to read. Note, that although diagrams for DRCR and SRLG-disjoint DCR problems look similar, the meaning of bars, colored in the same color is different for the two problems.
	
	We decided not to plot distribution for DRCR problem on real-world graphs, since those instances are too easy to solve for both state-of-the-art and our algorithms, and thus are not interesting. The distribution for a real-world graph instance of SRLG-disjoint DCR problem we can see on Figure \ref{fig:SRLG_space}.
	
\begin{figure*}[!ht]
		\centering
		\includegraphics[width= 1 \textwidth]{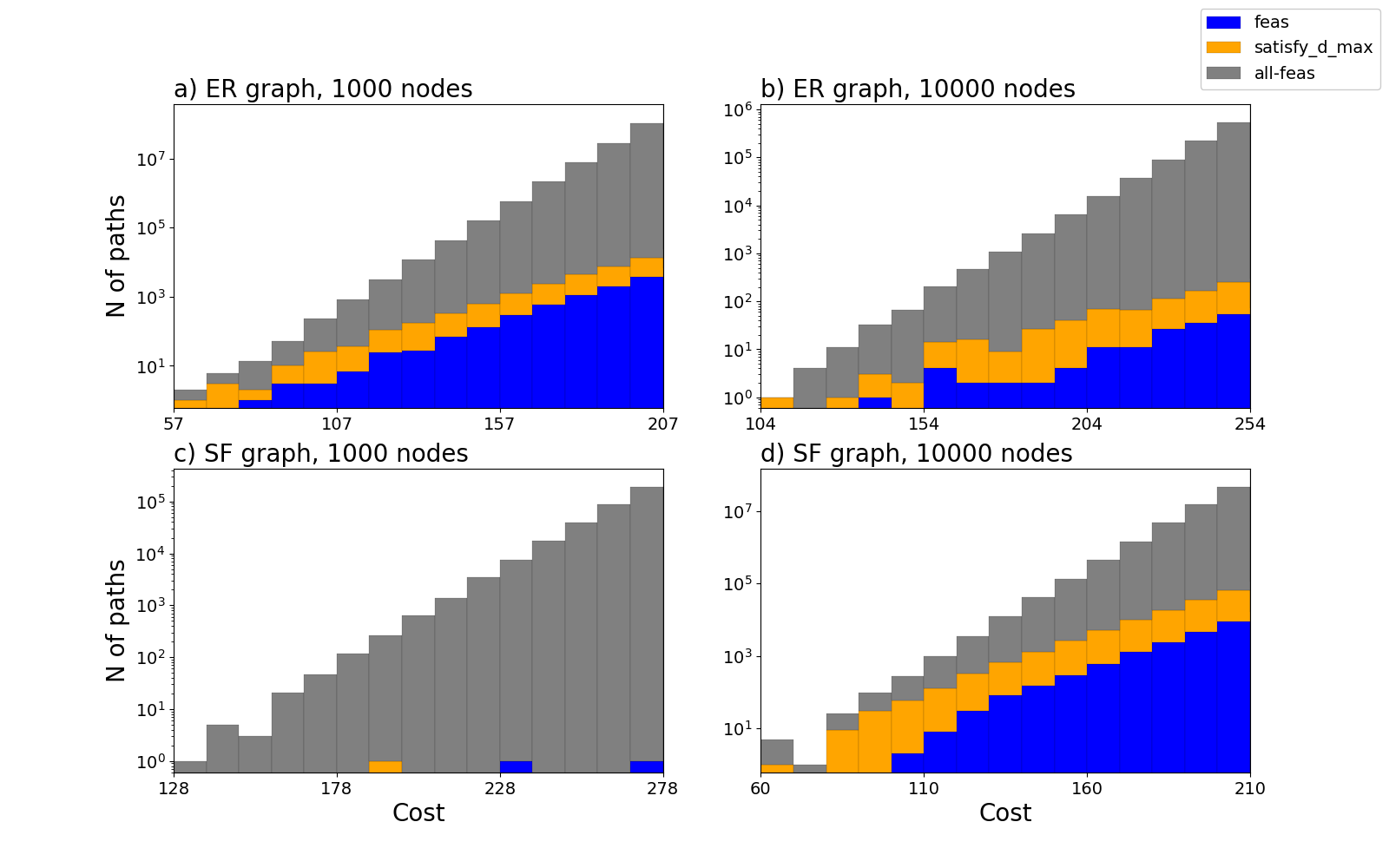}
		\caption{Low-cost tails of path distributions for DRCR problem. Randomly chosen instances of different size graphs of ER and SF types. Each bar shows number of paths, that have cost within the range of width 10. Blue paths are feasible, i. e. satisfying both upper and lower limit on delay. Orange paths are those satisfying only the upper limit. Grey paths are all paths, that connect source node to target node for the task. We used logarithmic scale on Y-axis in order to make picture more readable. From this fact one can see, that number of paths is growing almost exponentially at the lower end of the distribution, and the shortest feasible path is typically located not far from the shortest unconstrained one.}
		\label{fig:CSP_tails}
	\end{figure*}

\begin{figure*}[!ht]
		\centering
		\includegraphics[width=1 \textwidth]{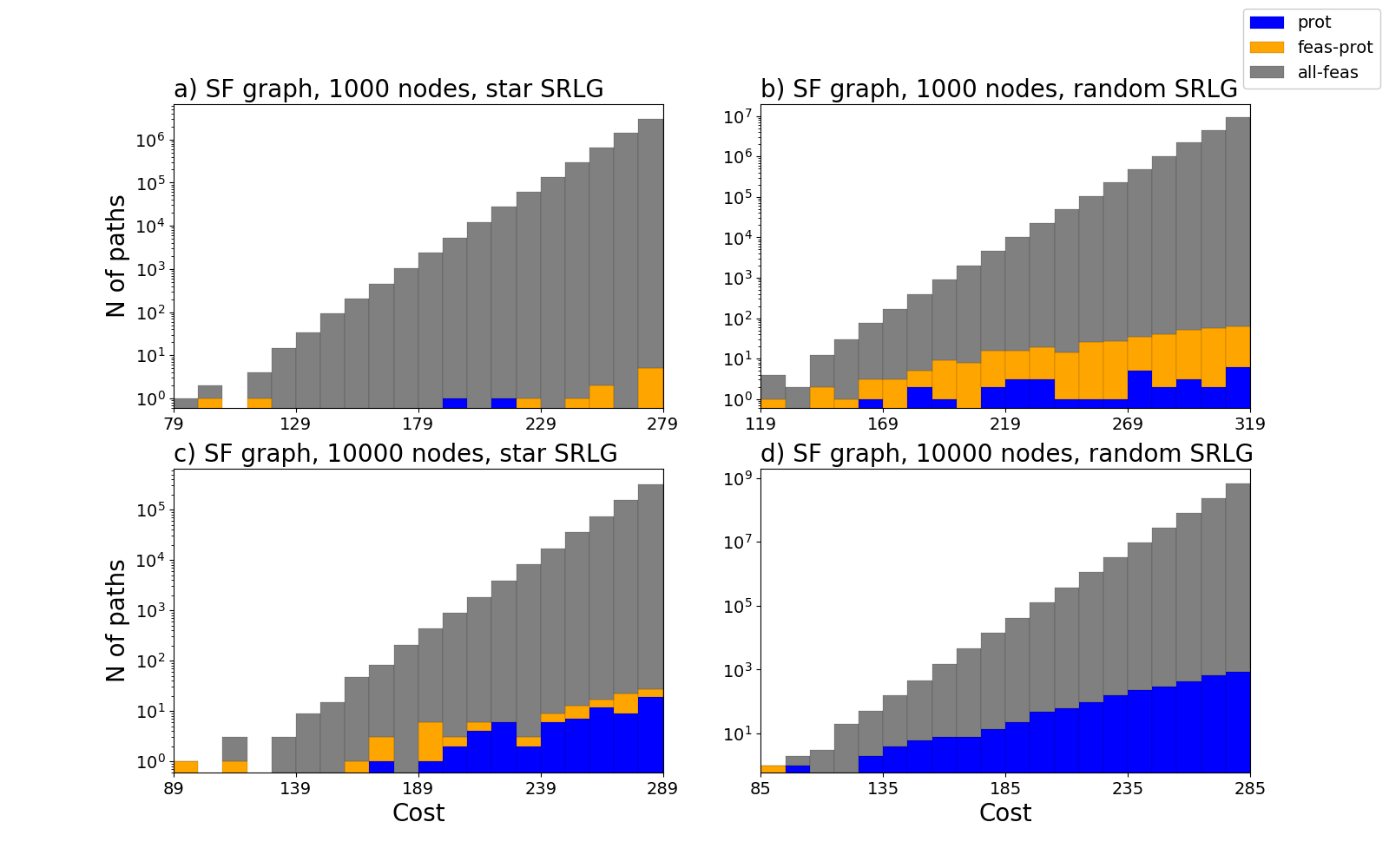}
		\caption{Low-cost tails of Active Path candidate distributions for SRLG-disjoint DCR problem for 4 instances of SF graphs. Both random and star SRLG generators are represented, as well as different graph sizes. The scale on Y-axis is logarithmic. Blue bars represent those AP candidates, that have PP. Orange bars are AP candidates, that satisfy constraints, but do not have disjoint PP pair. Grey bars are number of all paths in corresponding cost ranges.}
		\label{fig:SF_SRLG_tails}
	\end{figure*}

\begin{figure*}[!ht]
		\centering
		\includegraphics[width=1 \textwidth]{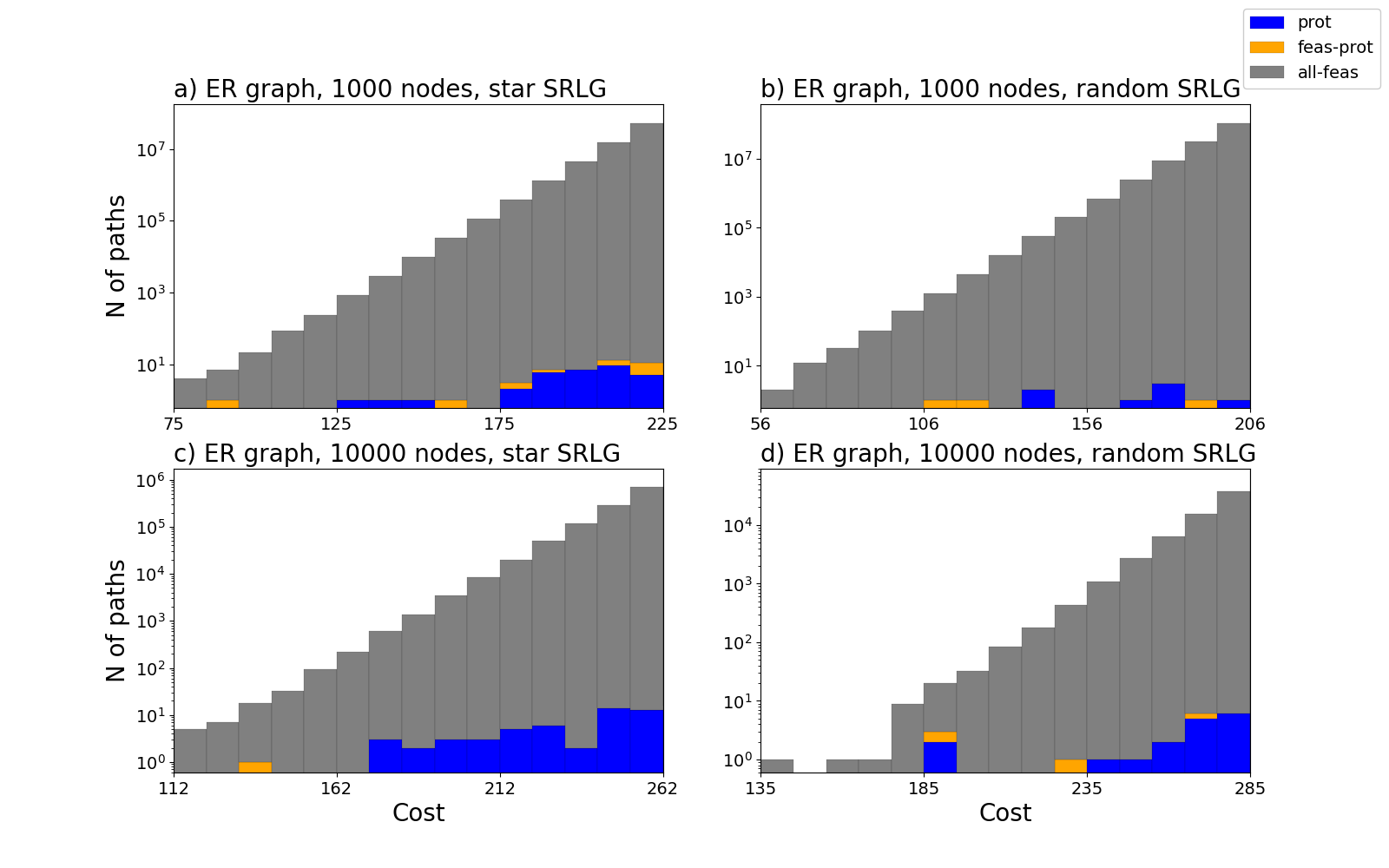}
		\caption{Low-cost tails of Active Path candidate distributions for SRLG-disjoint DCR problem for 4 instances of ER graphs. Both random and star SRLG generators are represented, as well as different graph sizes. The scale on Y-axis is logarithmic. Blue bars represent those AP candidates, that have PP. Orange bars are AP candidates, that satisfy constraints, but do not have disjoint PP pair. Grey bars are number of all paths in corresponding cost ranges.}
		\label{fig:ER_SRLG_tails}
	\end{figure*}

	It is important to note, that plots on figures \ref{fig:CSP_tails}, \ref{fig:SF_SRLG_tails} and \ref{fig:ER_SRLG_tails} are not intended to reflect distinctive features of each graph topology or SRLG generator type. Actually, even distributions for two different tasks on the same graph, i. e. two different pairs of source and target nodes and values of delay bounds may differ more, than those between graphs of different topologies. What is really important in the light of our work is that all of them have the optimal solution lying within the lower cost tail of the distribution, where feasible paths are rare. This general observation inspires us to use heuristic cost bounds and cost corridors in order to perform search only within the tail, keeping the search space small.
	
\begin{figure*}[!ht]
		\centering
		\includegraphics[width=1 \textwidth]{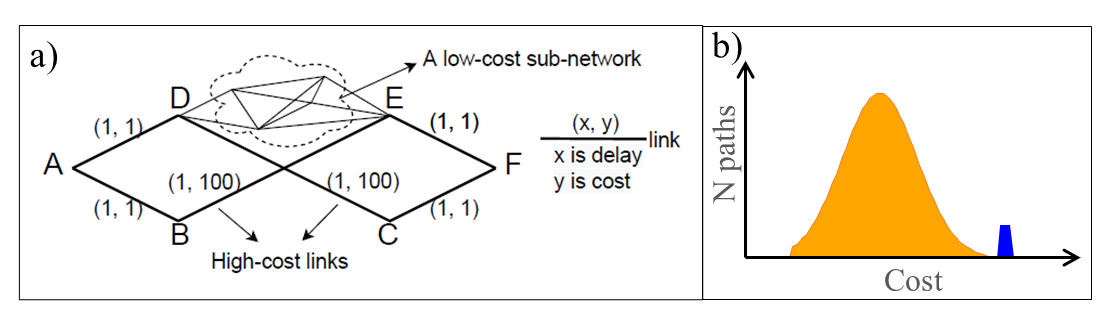}
		\caption{a) An example of graph topology, that falls into a trap issue, when solving SRLG-disjoint routing problem \cite{Zhao:23}. Despite being a clear case for explanatory purposes, this artificial topology would have path distribution significantly different from what we observe in real life scenarios. b) Expected path distribution with respect to cost for the shown graph topology. Depicted in orange are feasible AP candidates, that do not have any disjoint PP. The only two AP candidates, that have disjoint protection, lie at high cost values. This distribution is essentially different from what we see on real cases.}
		\label{fig:Trap_artificial_topology}
	\end{figure*}

	Finally, on the Figure \ref{fig:Trap_artificial_topology} we show, how the path distribution may look like for the example of trap topology, that is usually provided in literature. We note, that although it is good for illustrative purposes and allows clear demonstration of the notion of trap, it also represents an extreme case and may be misleading, if used for the discussion on real networks. It is widely known, that many NP-hard problems like the disjoint routing problem have extremely difficult worst case instances, but sometimes efficiently solvable on average in the real-life scenarios. So, practical approach should be concentrated on the features of real-life problems instead of the extreme artificial cases.
	
	\clearpage
	
	\section{Dependency of BTCS execution time on the search space parameters and graph features}
	\label{app:worst_case}
	
	In this section we present more data on how worst-case performance of BTCS algorithm is related to the path distribution and graph features. It is not intended to be an exhaustive analysis, but we believe that it gives useful perspective into the algorithm's nature. Note that only tasks that have feasible solution were selected for this experiment.
	
	\begin{figure*}[!ht]
		\centering
		\includegraphics[width= 1 \textwidth]{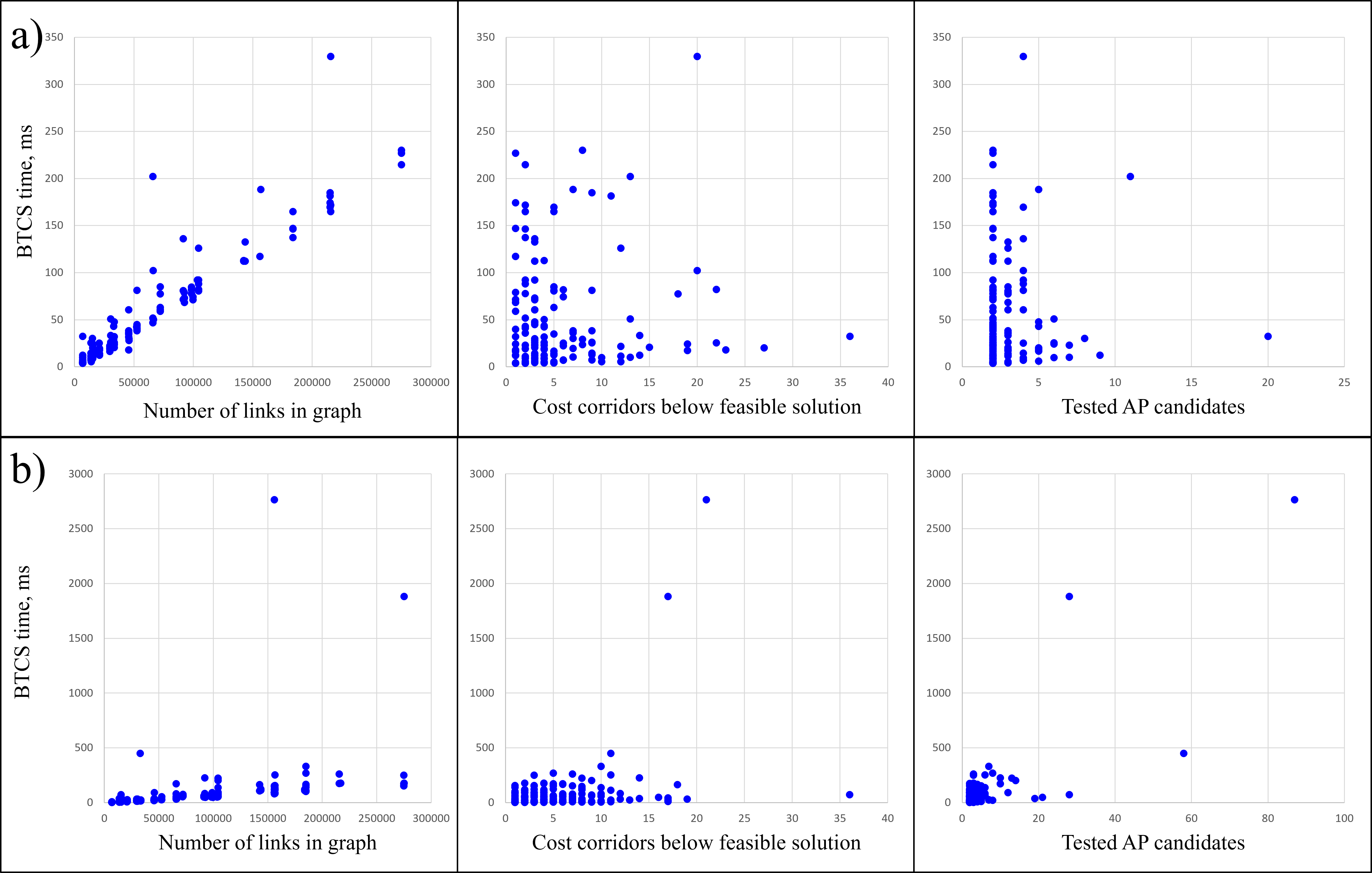}
		\caption{BTCS execution time dependency on parameters of the graph and the lower cost tail of AP search space. a) ER graphs with random SRLG, b) ER graphs with star SRLG.}
		\label{fig:ER_supp}
	\end{figure*}	
	
	\begin{figure*}[!ht]
		\centering
		\includegraphics[width= 1 \textwidth]{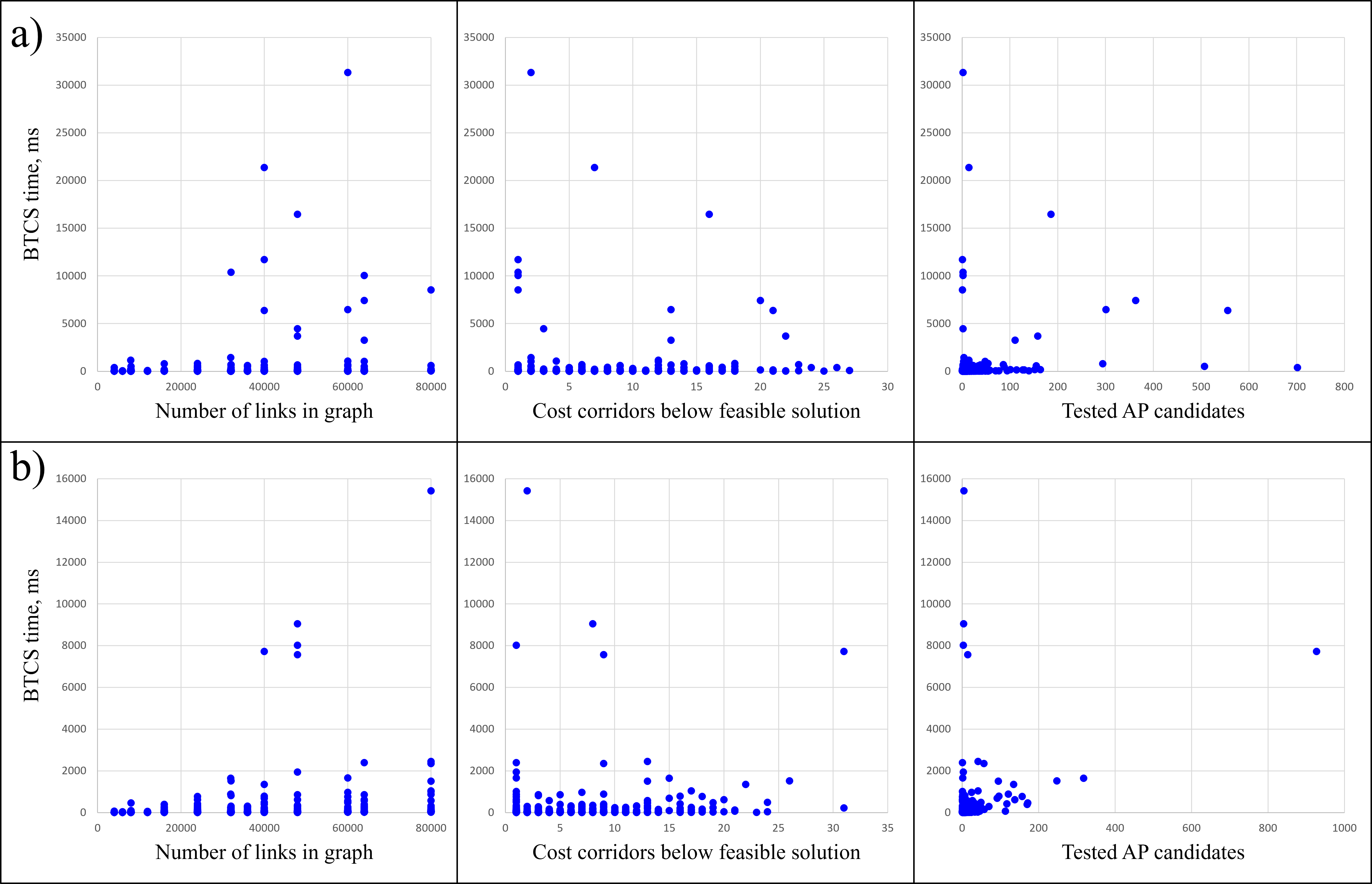}
		\caption{BTCS execution time dependency on parameters of the graph and the lower cost tail of AP search space. a) SF graphs with random SRLG, b) SF graphs with star SRLG.}
		\label{fig:SF_supp}
	\end{figure*}	
	
\end{appendices}
\vfill

\end{document}